\documentclass[aps,prb,twocolumn,showpacs,superscriptaddress,
groupedaddress,10pt]{revtex4}
\usepackage{amsmath}
\usepackage{amssymb}
\usepackage{graphicx}
\usepackage{epsfig}
\usepackage{dcolumn}
\usepackage{bm}
\usepackage{bm}
\usepackage{blindtext}
\usepackage{natbib}

\usepackage{mathtools}

\DeclarePairedDelimiter\floor{\lfloor}{\rfloor}
\def\avg#1{\langle#1\rangle}
\def\Re{\rm{Re}}

\def\be{\begin{equation}} \def\ee{\end{equation}}
\def\bea{\begin{eqnarray}} \def\eea{\end{eqnarray}}

\def\nn{\nonumber}
\def\pp{\parallel}

\newcommand{\ket}[1]{| #1 \rangle}
\newcommand{\bra}[1]{\langle #1 |}

\begin{document}
\title{
One-dimensional Quantum Spin Dynamics of Bethe String States
}
\author{Wang Yang}
\affiliation{Department of Physics, University of California,
San Diego, California 92093, USA}
\author{Jianda Wu}
\email{wjdandeinstein@gmail.com}
\affiliation{Department of Physics, University of California,
San Diego, California 92093, USA}
\author{Shenglong Xu}
\affiliation{Department of Physics, University of California,
San Diego, California 92093, USA}
\author{Zhe Wang}
\affiliation{Experimental Physics V, Center for Electronic Correlations
and Magnetism, Institute of Physics, University of Augsburg, 86135
Augsburg, Germany}
\author{Congjun Wu}
\email{wucj@physics.ucsd.edu}
\affiliation{Department of Physics, University of California,
San Diego, California 92093, USA}

\begin{abstract}
Quantum dynamics of strongly correlated systems is a challenging problem.
Although the low energy fractional excitations of one dimensional
integrable models are often well-understood, exploring quantum dynamics
in these systems remains challenging in the gapless regime, especially at  intermediate and high energies.
Based on the algebraic Bethe ansatz formalism, we study spin dynamics
in a representative one dimensional strongly correlated model, {\it i.e.},
the antiferromagnetic spin-$\frac{1}{2}$ XXZ chain with the Ising
anisotropy, via the form-factor formulae.
Various excitations at different energy scales are identified crucial
to the dynamic spin structure factors under the guidance of sum rules.
At small magnetic polarizations, gapless excitations dominate
the low energy spin dynamics arising
from the magnetic-field-induced incommensurability.
In contrast, spin dynamics at intermediate and high energies is
characterized by the two- and three-string states,
which are multi-particle excitations based on the commensurate
N\'eel ordered background.
Our work is helpful for experimental studies on spin dynamics
in both condensed matter and cold atom systems
beyond the low energy effective Luttinger liquid theory.
Based on an intuitive physical picture,
we speculate that the dynamic feature at high energies due
to the multi-particle anti-bound state excitations can be
generalized to non-integrable spin systems.
\end{abstract}
\maketitle

\section{Introduction.}
The real-time dynamics reveals rich information of the quantum nature
of strongly correlated many-body states \cite{Deift1994,Sagi1996,Sachdev1997,
Konik2003,Coldea2010,Sachdev2011,Ganahl2012,Imambekov2012,Fukuhara2013,
Wu2014,Caux2016,Aronson2016,Babadi2015,andrei2016}.
On the other hand, one-dimensional integrable models due to their
exact solvability provide reliable reference points for studying
quantum and thermodynamic correlations \cite{Bethe1931,Yang1966,
Yang1966a,Faddeev1979,Baxter1982,Takahashi2005,Guan2013,Wang2015},
and certain characteristic features exhibited in these integrable models
are relevant to even non-integrable systems.
The spin-$\frac{1}{2}$ antiferromagnetic (AFM) Heisenberg XXZ chain,
a representative of integrable models, is an ideal system for
a non-perturbative study on quantum spin dynamics
\cite{Karbach2000,Biegel2002,Sato2004,Caux2005,Caux2005a,Caux2009,
Kohno2009,Mossel2010,Liu2014,ishimura1980}.
Nevertheless, it remains a very challenging problem due to the interplay
between quantum fluctuations and the dynamic evolution.
On the experimental side,
a great deal of high precision measurements have been
performed on quasi one-dimensional ($1$D) materials by using
neutron scattering and electron spin resonance (ESR) spectroscopy
\cite{yoshizawa1981, Nagler1982, Nagler1983, Nagler1991,Zheludev2000,Stone2003,He2005,Kimura2007,
Mourigal2013,Wang2015a,Aronson2016}.
These systems are faithfully described by the 1D spin-$\frac{1}{2}$ AFM
Heisenberg model.

There has appeared significant progress in calculating the
dynamic spin structure factors (DSSF)
\cite{Karbach2000,Biegel2002,Sato2004,Caux2005,Caux2005a,
Caux2009,Kohno2009,Mossel2010,Liu2014}.
At zero field, contributions to the DSSFs from the two- and
four-spinon excitations can be solved analytically
by using the quantum affine symmetry
\cite{Bougourzi1996,Abada1997,Bougourzi1998,Caux2012,Jimbo1994},
however, this method ceases to apply at nonzero fields.
In the algebraic Bethe ansatz formalism \cite{Faddeev1979,Korepin1997},
the matrix elements of local spin operators between two different
Bethe eigenstates are expressed in terms of the determinant formulae
in finite systems \cite{Korepin1982,Slavnov1989,Maillet2000,Kitanine2000}.
Accompanied with a judicious identification of the dominant excitations
to spin dynamics,
this method can be used to efficiently calculate the DSSFs for considerably large systems.
Excellent agreements between theories and experiments have been established
for the SU(2) invariant spin-$\frac{1}{2}$ AFM Heisenberg chain,
confirming the important role of spinon excitations
in the dynamic properties \cite{Mourigal2013}.

In this article, we study quantum spin dynamics in an AFM spin-$\frac{1}{2}$ XXZ chain with the Ising anisotropy at zero temperature in a longitudinal
magnetic field.
The spin chain under consideration is gapped at zero field, and
increasing field tunes the system into the gapless regime \cite{Yang1966a},
in which the full spin dynamics remains to be explored.
Working within the algebraic Bethe ansatz formalism, we identify
various spin excitations separated at different energy scales.
The $S^{-+}(q,\omega)$-channel is dominated by the psinon pair
excitations resembling the zero field des Cloizeaux-Pearson
(DCP) modes \cite{DesCloizeaux1962}, whose momentum range shrinks as increasing polarization.
The coherent low energy excitations of the $S^{+-}(q,\omega)$ resemble
the Larmor mode at $q\to 0$, and become incoherent at $q\to \pi$.
The 2- and 3-string states play important roles at intermediate
and high energies, reflecting the background N\'eel configuration.
The low energy excitations in the longitudinal $S^{zz}(q,\omega)$
channel exhibit the sound-like spectra at $q\to 0$ while the spectra
in the high energy sector reflect the excitonic excitations
on the gapped N\'eel background.
These high-frequency features of spin dynamics cannot
be captured by the low energy effective Luttinger liquid theory.
Based on a simple physical picture, we argue that the revealed dynamic features are also relevant to non-integrable cases.

The rest part of this article is organized as follows.
In Sect. \ref{sect:ham}, the model Hamiltonian is presented.
In Sect. \ref{sect:method}, the method of algebraic Bethe ansatz
and the calculation method are introduced.
In Sect. \ref{sect:transverse}, the results of the
transverse DSSFs are calculated.
In Sect. \ref{sect:longitudinal}, the results of the
longitudinal DSSFs are calculated.
Discussions and conclusions are made in Sect.
\ref{sect:discussion}.
Various details of calculations are presented in
Appendices A - F.

\section{The model Hamiltonian}
\label{sect:ham}

The Hamiltonian of the 1D spin-$\frac{1}{2}$ AFM chain with the periodic
boundary condition in the longitudinal magnetic field $h$ is defined as
\bea
H_{0}&=&J\sum_{n=1}^{N} \left \{ S^{x}_{n} S^{x}_{n+1} + S^{y}_{n} S^{y}_{n+1}
+\Delta \left(S^{z}_{n} S^{z}_{n+1}-\frac{1}{4}\right)\right \}, \nn \\
H&=& H_{0}- h \sum_{n=1}^{N} S^{z}_{n}
\label{eq:Hamiltonian},
\eea
where $N$ is the total site number.
The spin operators on the $n$-th site are $S_n^{\alpha}=\frac{1}{2}\sigma^\alpha$ with $\alpha=x,y,z$.
We consider the axial region with the anisotropic parameter
$\Delta=\cosh\eta>1$.

The ground state at zero field is known to exhibit the long-range
Neel ordering, and, hence, is spin gapped.
If the external field $h$ is small, then there is no magnetization.
The magnetization $m= \avg{G|S^z_T|G}/N$ starts to develop when $h$
is above a critical value $h_c(\Delta)$, and then the system
enters the gapless regime, where $|G\rangle$ represents
the ground state and $S^z_T=\sum_{i=1}^N S^z_i$ is the $z$-component
of total spin.
$h$ and $m$ are conjugate variables through the relation
$h={\partial e_{0}}/{\partial m}$ with $e_0=\avg{G|H_0|G}/N$.
For calculations presented below, we adopt a typical
value of $\Delta=2$ (which applies to the $\text{SrCo}_{2}\text{V}_{2}\text{O}_{8}$ material \cite{Wang2018}) and
$N=200$ unless explicitly mentioned, and the
corresponding critical field is $h_c/J=0.39$ \cite{Yang1966a}.

We will calculate the zero temperature DSSFs, which are expressed
in the Lehman representation as
\bea
S^{a\bar{a}}(q,\omega)
&=& 2\pi \sum_{\mu} |\bra{\mu} S^{\bar{a}}_{q} \ket{G}|^{2}
\delta(\omega-E_{\mu}+E_{G}), \ \ \,
\label{eq:lehmann}
\eea
where $a=\pm$ and $z$; $\bar a=-a$ for $a=\pm$, and $a=\bar a$
for $a=z$;
$S_i^\pm=\frac{1}{\sqrt{2}}(S_x\pm i S_y)$
and the Fourier component of spin is defined as
\bea
S^{a}_{q}= \frac{1}{\sqrt N} \sum_{j} e^{iqj} S^{a}_{j};
\eea
$\ket{\mu}$ is the complete set of eigenstates;
$E_{G}$ and $E_{\mu}$ are eigenenergies of the ground
and excited states, respectively.

\section{The Bethe ansatz method}
\label{sect:method}
In this section, we briefly describe the Bethe ansatz method that we
employ to calculate the DSSF.
The fully polarized state with all spins up is taken as the
reference state, based on which the flipped spins are viewed as particles.
A state with $M$ flipped spins is denoted an $M$-particle state
and the polarization $m=1/2-M/N$.
Each particle wavevector $k_j$ is related to a rapidity $\lambda_j$
through the relation
\bea
e^{ik_j}=\sin(\lambda_j
+i\frac{\eta}{2})/\sin(\lambda_j-i\frac{\eta}{2}).
\eea
The set of rapidities $\{\lambda_j\}$ with ($1\le j\le M$) are
determined by the integer or half-integer-valued Bethe quantum numbers $I_j$
as presented in 
Appendix \ref{sect:BA}.
The ``psinon''-pair states $n\psi\psi$ and ``psinon-antipsinon" pair states $n\psi\psi^*$
($n=1,2$) with $n$ the pair number
play important roles in both transverse and longitudinal DSSFs.
These eigenstates possess real rapidities \cite{Karbach2002,Caux2005a}
and their Bethe quantum numbers are presented in 
Appendix \ref{sect:BA}.

If some $\lambda_j$'s are complex \cite{Bethe1931}, the
corresponding states are termed as string states \cite{Takahashi2005}
in which some particles form bounded excitations
as discussed in
Appendix \ref{sect:string}.
The string ansatz is an approximation
assuming the string pattern of the complex rapidity distribution.
A length-$l$ ($l\ge 1$) string is denoted as $\chi^{(l)}$, which represents
a set of complex rapidities
\bea
\lambda^{(l)}_{j}= \lambda^{(l)}+i\frac{\eta}{2}(l+1-2j),
\eea
for $1\le j\le l$.
Their common real part $\lambda^{(n)}$, the string center, is determined
from the Bethe-Gaudin-Takahashi (BGT) equations with the reduced Bethe
quantum numbers \cite{Takahashi2005} shown in 
Appendix \ref{sect:string}.

Below we only consider the solutions with one length-$l$ string denoted
as $1\chi^{(l)}R$ where $R=m\psi\psi^*$ or $m\psi\psi$.
The errors of complex rapidities are used to judge the validity of
the string ansatz, which can be analytically checked \cite{Hagemans2007}.
For the calculated range of $2m$ from 0.1 to 0.9, our results
exhibit a high numeric accuracy.
A bar of $10^{-6}$ is set and only string states within this bar are kept
in calculating DSSFs.
The detailed discussions on the error estimation and how to
systematically improve the string ansatz in an exact manner
are included in 
Appendix \ref{sect:deviation}.

The determinant formulae for the form factors $\bra{\mu} S^{\pm}_{j} \ket{G}$
can be obtained from the rapidities as presented in
Ref. [\onlinecite{Kitanine2000}]
and as summarized in 
Appendix \ref{sect:determinant}.
Due to the exponentially large number of excited states,
only a subset of them with dominating contributions to the DSSFs
are selected.
The validity of the selection is checked by comparing the results
with the exact sum rules,
and these sum rules are derived in Appendix \ref{sect:sum}.

\section{The transverse Dynamic spin structure factor}
\label{sect:transverse}

In this section, we discuss the dominant contributions of excited
states to the transverse DSSFs include $n\psi\psi^* (n=1,2)$,
$1\chi^{(2)}R$ and $1\chi^{(3)}R$ where $R=1\psi\psi^*$, and $1\psi\psi$.
We also check the saturation of these excitations by comparing
with the exact sum rules.

\subsection{The momentum-resolved sum rule of the transverse DSSF}
\label{sect:sum_resolved}

\begin{figure}
\centering\epsfig{file=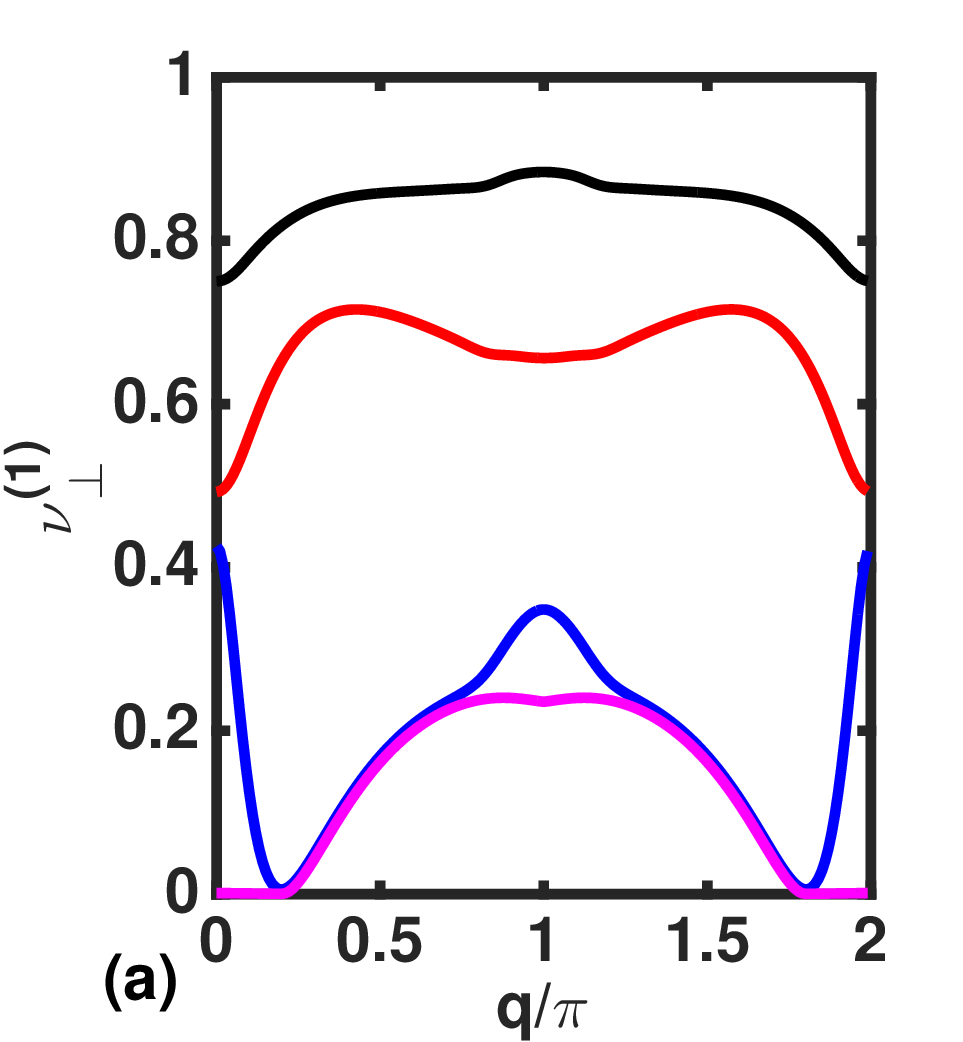,clip=0.4,
width=0.35\linewidth,height=0.4\linewidth,angle=0}
\centering\epsfig{file=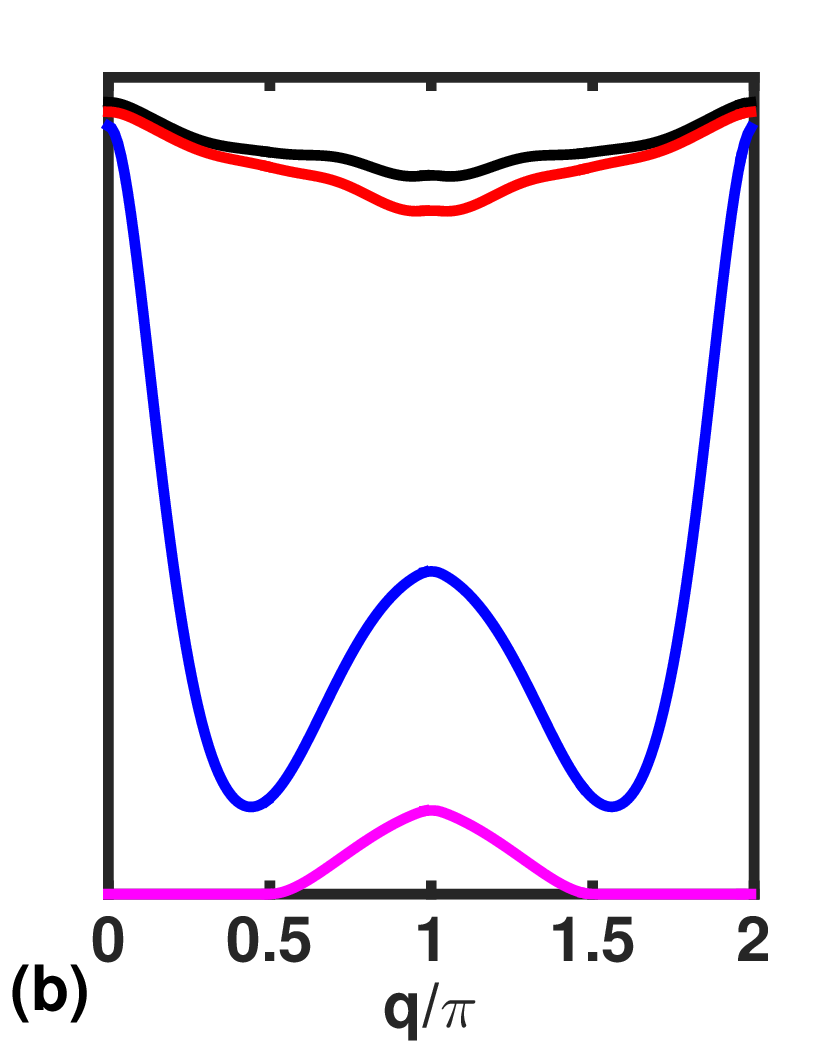,clip=0.4,
width=0.30\linewidth,height=0.4\linewidth,angle=0}
\centering\epsfig{file=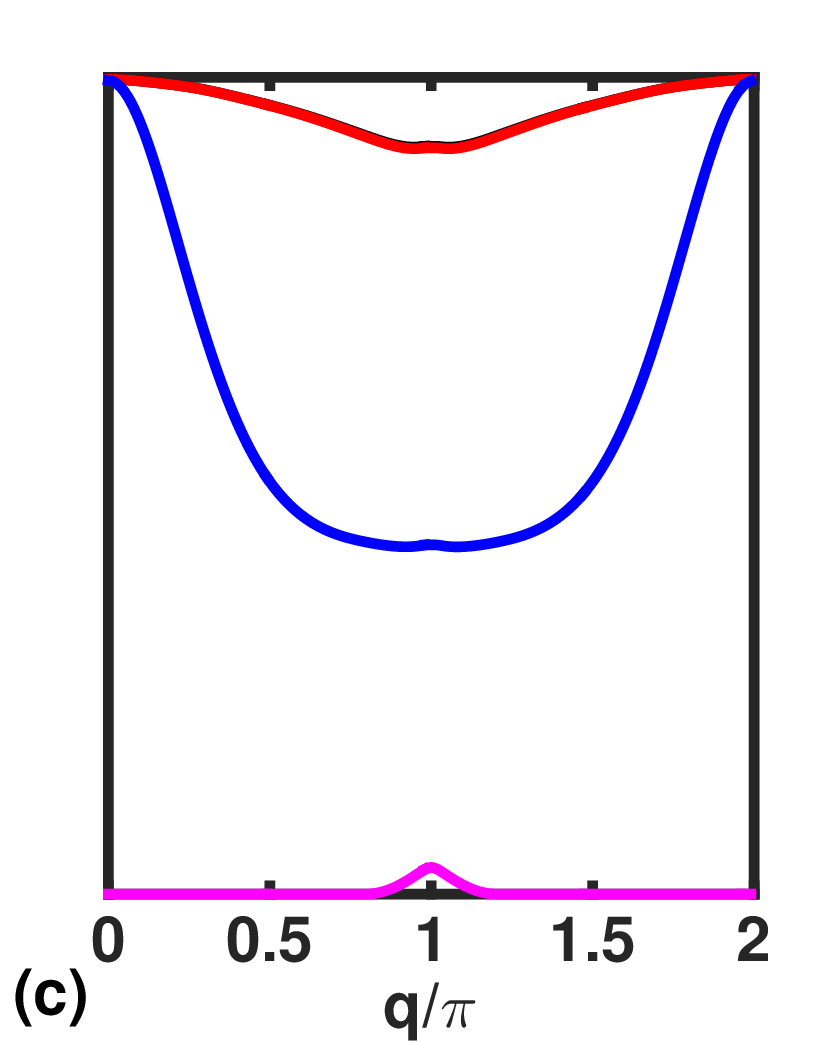,clip=0.4,
width=0.30\linewidth,height=0.4\linewidth,angle=0}
\caption{The momentum-resolved FFM ratios with $2m$ equal to ($a$) $0.2$,
($b$) $0.5$, and ($c$) $0.8$, respectively.
The pink, blue, red and black curves represent cumulative results
by including the psinon states $n\psi\psi$ ($n=1,2$) in $S^{-+}$,
the psinon-antipsinon states $n\psi\psi^*$ ($n=1,2$), the
2-string states and 3-string states in $S^{+-}$, respectively.
In ($a$), the pink and blue curves overlap significantly
and so do the red and black curves in ($c$).
}
\label{fig:ratio}
\end{figure}

The transverse first frequency moment (FFM) sum rule is
\bea
W_\perp(q) &=& \int_{0}^\infty  \frac{d \omega}{2\pi}~
\omega \left[ {S^{ +  -} (q,\omega ) + S^{ -  + } (q,\omega )} \right] \nn \\
&= &  \alpha_\perp +\beta_\perp \cos q ,
\label{eq:ffm}
\eea
where $\alpha_\perp= -e_0-\Delta{\partial e_0}/{\partial \Delta}+mh$
and $\beta_\perp =(2-\Delta ^2){\partial e_0}/{\partial \Delta}
+\Delta e_0$.
To evaluate the saturation levels,
we define the ratio of the momentum-resolved FFMs as
\bea
\nu^{(1)}_{\perp}(q)=\tilde{W}_{\perp}(q)/W_{\perp}(q),
\eea
where $\tilde{W}_{\perp}(q)$ is calculated from
the partial summations over the selected excitations.

The calculated momentum-resolved transverse FFM ratios
$\nu^{(1)}_{\perp}(q)$ in the Brillouin zone are displayed in Fig.~\ref{fig:ratio} for three representative magnetizations
of $m=0.2, 0.5$, and $0.8$.
The magnetic polarization breaks time-reversal symmetry, and thus
$S^{+-}$ contributes more prominently than $S^{-+}$ to sum rules.
We start with plotting $S^{-+}$ contributions, which take
into account the ``psinon''-pair states $n\psi\psi$
($n=1,2$) with $n$ the pair number.
These eigenstates possess real rapidities \cite{Karbach2002,Caux2005a}
and their Bethe quantum numbers are presented in 
Appendix \ref{sect:BA}.

The $S^{+-}$ channel is more involved: Dominant excitations include
the ``psinon-antipsinon'' pair states denoted as $n\psi\psi^*$
and string states.
Combined with $S^{-+}$, different contributions are plotted
and their relative weights are displayed explicitly.
The $n\psi\psi^*$ excitations are with real rapidities and their
Bethe quantum numbers are given in Appendix \ref{sect:BA}.
These states with $n=1$ and $2$ contribute significantly to $S^{+-}(q,\omega)$
at high polarizations, particularly at long wave lengths.
But their weights become less important as decreasing polarization.
This observation is supported by considering the limit of $2m\to 0$
at $S_{T}^z=1$, then $|\mu\rangle$'s in Eq. \ref{eq:lehmann} belong
to the subspace of $S^z_T=0$, whose dimension is $N!/(\frac{N}{2}!)^2$.
In this sector, there only exist two states with all real
rapidities representing even and odd superpositions of
two symmetry breaking N\'eel states.
The dominant weights near the critical line $h_c(\Delta)$ should arise
from string states.

The calculation for $S^{+-}(q,\omega)$ is significantly improved by
including the string state contributions shown in Fig.~\ref{fig:ratio}.
The two-string excitations $1\chi^{(2)}R$ $(R=1\psi\psi^*,
1\psi\psi)$ greatly improves the saturation level of the FFM ratios
for both intermediate and high polarizations at all momenta.
In particular, the $1\chi^{(2)}1\psi\psi^{*}$ contributions are more
dominant than $1\chi^{(2)}1\psi\psi$, typically one order higher.
However, at small polarizations,  the two-string contributions decrease
quickly in particular at long wavelengths, indicating the
necessity of including states with even longer strings.
Including the 3-string excitations $1\chi^{(3)}1\psi\psi^*$
further improves the saturation level of $\nu^{(1)}_\perp(q)$
at small polarizations, while their contributions are minor
above the half-polarization.
The $1\chi^{(3)}1\psi\psi$ excitations are neglected since
their contributions are about two orders smaller.
After combining all the excitations above, a high saturation
level ($>80\%$) is reached for all momenta at the intermediate
(e.g. $2m=0.5$) and high polarizations (e.g. $2m=0.8$).
At small polarizations (e.g. $2m=0.2$), $\nu^{(1)}(q)$ is still
well saturated for most momenta.
Nevertheless, the saturation level decreases when $m \to 0$ at $q = 0$,
and the trend is more prominent for even smaller polarization.
There may exist unknown modes with significant weights around zero momentum.

\subsection{String states and spin dynamics}

\begin{figure}
\centering\epsfig{file=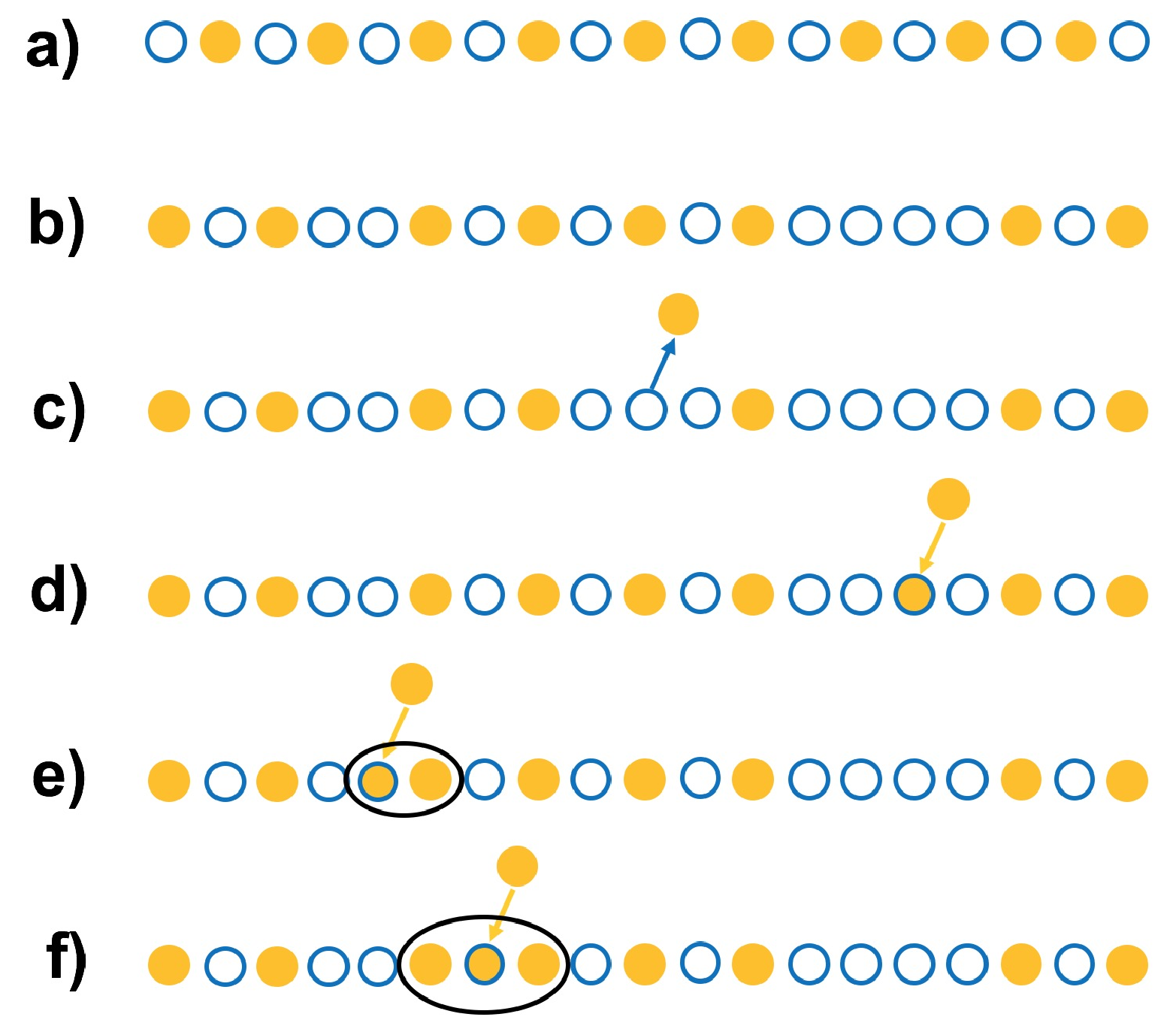,clip=0.5
width=0.7\linewidth,height=0.7\linewidth,angle=0}
\caption{Schematic plot of a representative spin configuration in the real space within:
$a)$ the N\'eel ordered ground state at zero field;
$b)$ the incommensurate ground state at a nonzero field $h>h_c$;
$c)$ a state with real particle wavevectors contributing to $S^{-+}$;
$d)$ a state with real particle wavevectors contributing to $S^{+-}$;
$e)$ a 2-string state contributing to $S^{+-}$;
$f)$ a 3-string state contributing to $S^{+-}$.
The blue hollow circle represents a spin up which is viewed as vacuum,
and the yellow solid circle represents a spin down which is viewed as a particle.
A particle is removed from (added to) the incommensurate ground state configuration in $S^{-+}$ ($S^{+-}$),
which is represented by an arrow pointing out of (into) the corresponding position in $b)$.
}
\label{fig:melting}
\end{figure}

The appearance of string states can be inferred based on an intuitive physical picture.
Fig. \ref{fig:melting} $a)$ shows a pictorial plot of a representative spin configuration in the N\'eel ordered ground state at zero field.
The system becomes incommensurate at $h>h_c$ as shown in Fig. \ref{fig:melting} $b)$,
but there is still a reminisce of the N\'eel ordering when the magnetization is small.
The excited states contributing to $S^{-+}$ have one less particle than the ground state.
As shown in Fig. \ref{fig:melting} $c)$, removing a particle leads to a configuration which still consists of unbound particles.
Hence the dominant excitations in $S^{-+}$ are Bethe eigenstates with real rapidities.

On the other hand, the states in $S^{+-}$ have one more particle than the ground state and the situation is more complicated with three possibilities.
If the particle is added into the region where the N\'eel ordering is absent,
all particles in the resulted excited state remain to be unbounded as shown in Fig. \ref{fig:melting} $d)$.
The second possibility is to bind the new particle with another existing particle, which gives a 2-string state displayed in Fig. \ref{fig:melting} $e)$.
Fig. \ref{fig:melting} $f)$ plots the third possibility of a 3-string state:
The additional particle can be inserted into the middle position of two particles and they form a three-body bounded entity.
Based on the above configuration of a diluted N\'eel ordering state,
adding a particle cannot create four particles in a row,
hence string states of higher orders occur with much rarer chances,
mainly as high order fluctuation effects.
Therefore, the $S^{+-}$ DSSF should be dominated by the above
three types of excited states.
We also expect that the roles played by string states will diminish as increasing the magnetic polarization, but are enhanced by increasing the anisotropy.
These intuitive considerations are supported by the Bethe ansatz calculations to be discussed below.

\subsection{The spectral weights}

\begin{figure}
\centering\epsfig{file=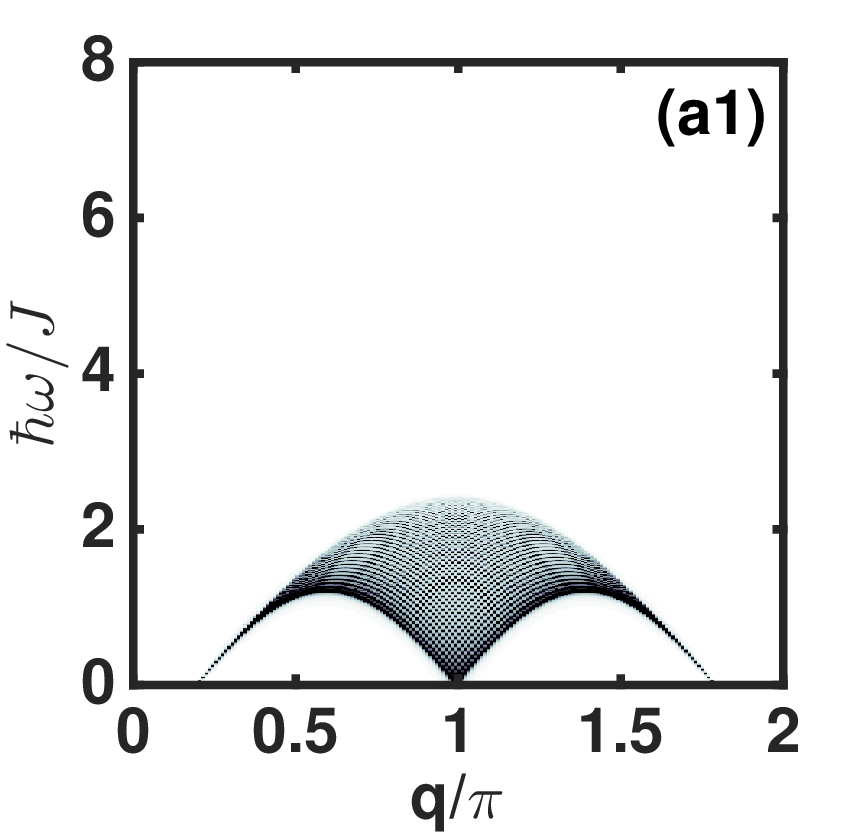,clip=0.4,
width=0.325\linewidth,height=0.32\linewidth,angle=0}
\centering\epsfig{file=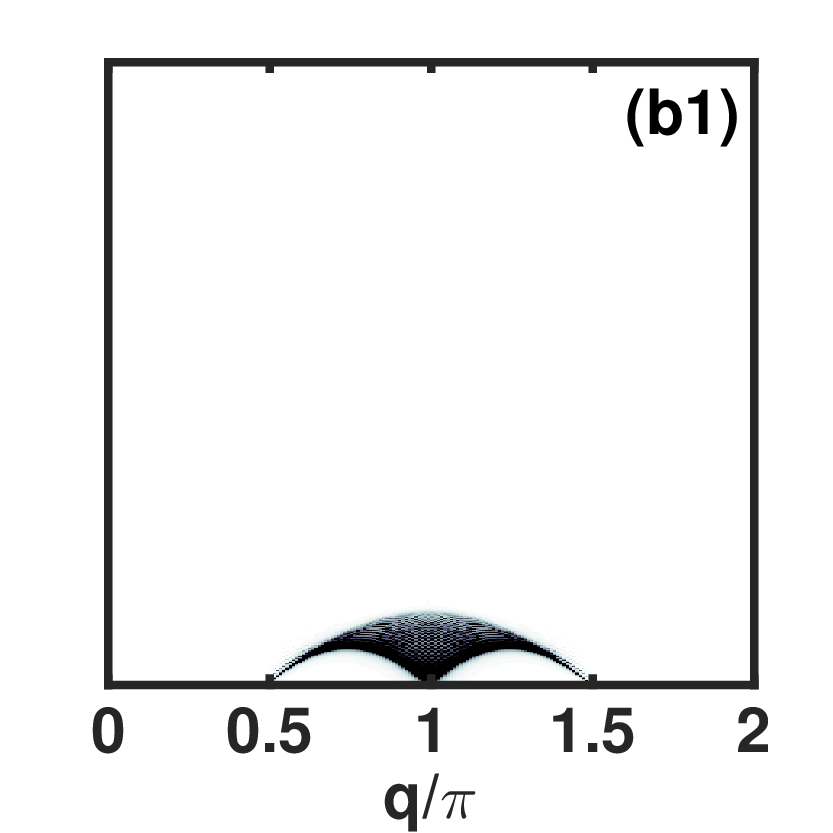,clip=0.4,
width=0.31\linewidth,height=0.32\linewidth,angle=0}
\centering\epsfig{file=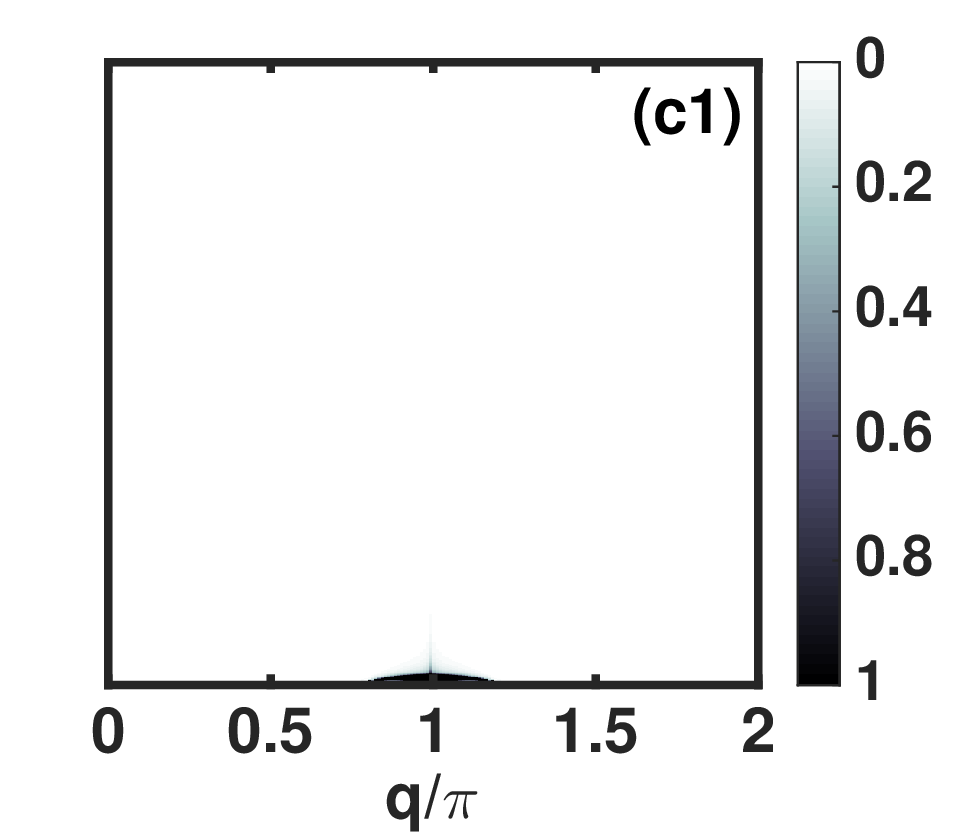,clip=0.4,
width=0.335\linewidth,height=0.32\linewidth,angle=0}
\centering\epsfig{file=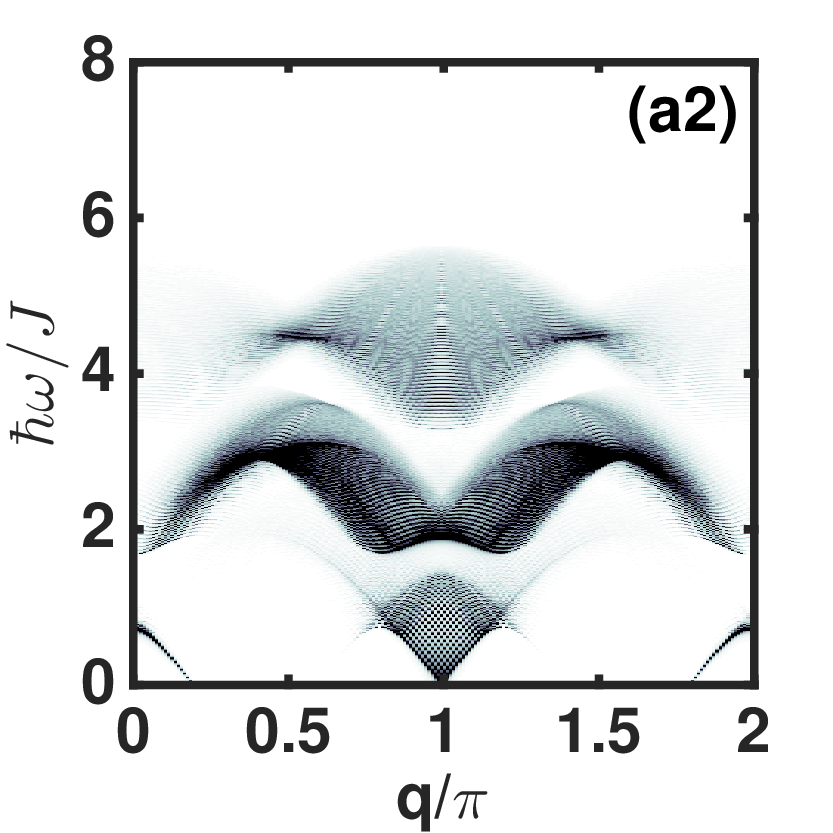,clip=,
width=0.32\linewidth,height=0.32\linewidth,angle=0}
\centering\epsfig{file=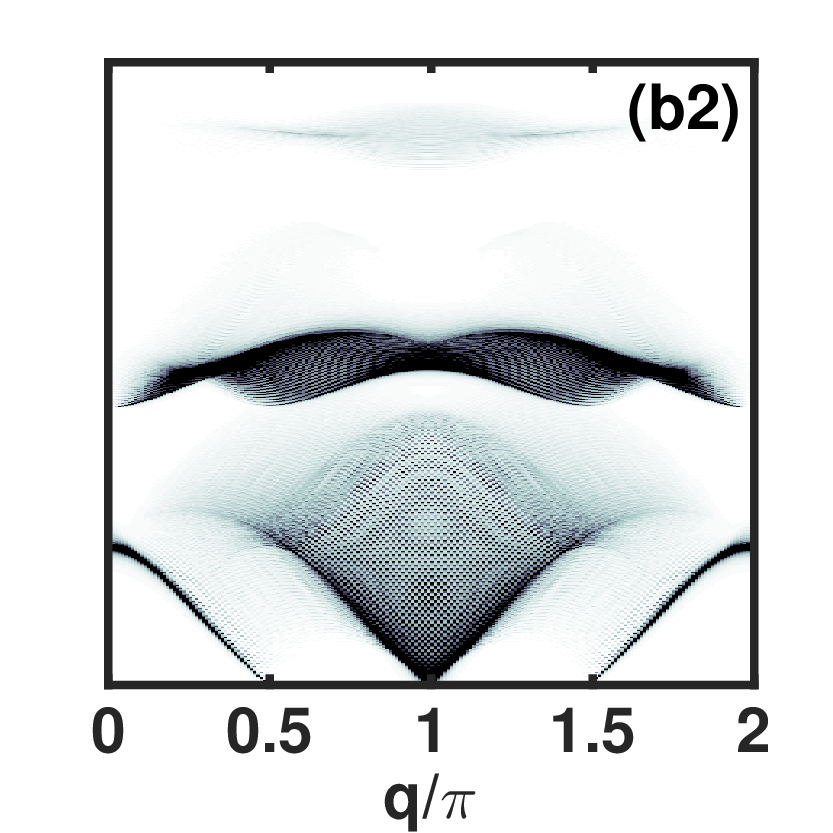,clip=0.4,
width=0.31\linewidth,height=0.32\linewidth,angle=0}
\centering\epsfig{file=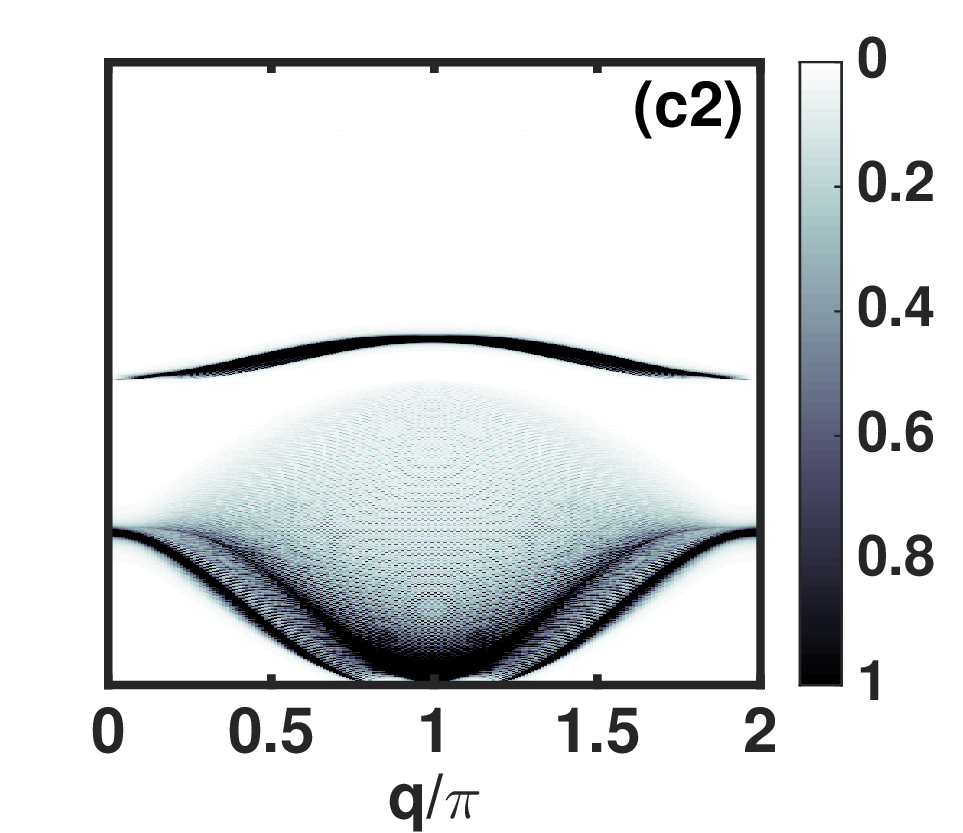,clip=0.4,
width=0.33\linewidth,height=0.32\linewidth,angle=0}
\centering\epsfig{file=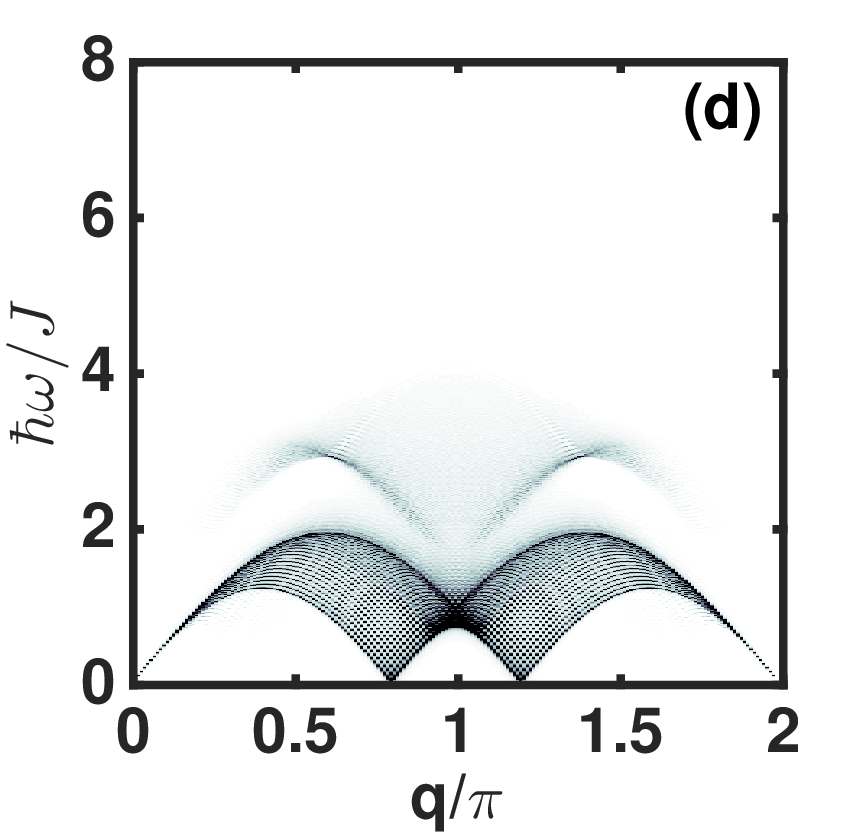,clip=0.4,
width=0.32\linewidth,height=0.32\linewidth,angle=0}
\centering\epsfig{file=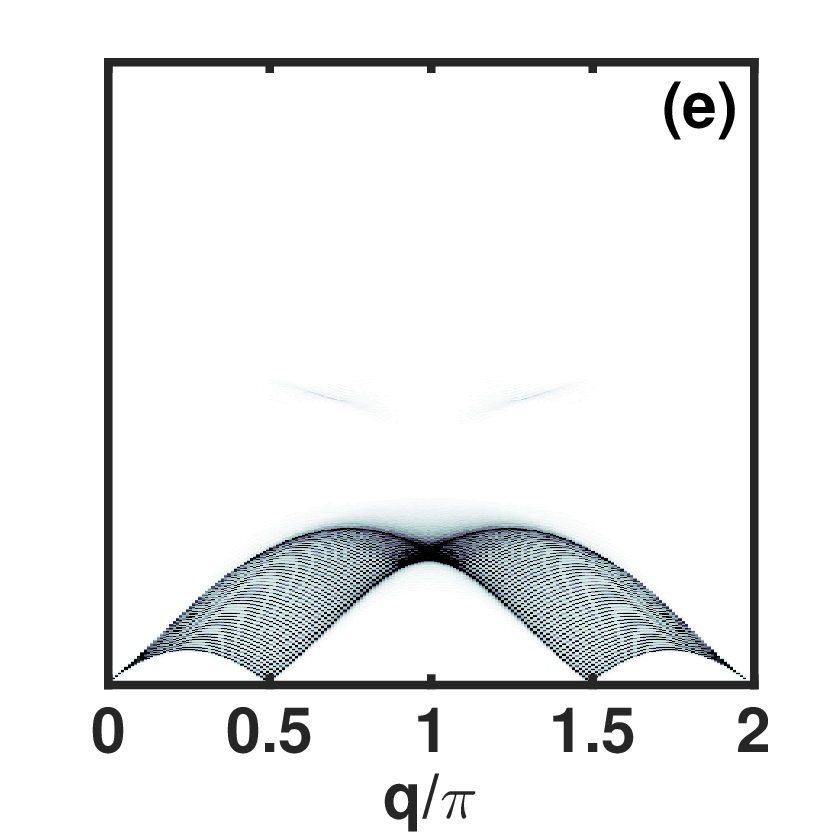,clip=0.4,
width=0.31\linewidth,height=0.32\linewidth,angle=0}
\centering\epsfig{file=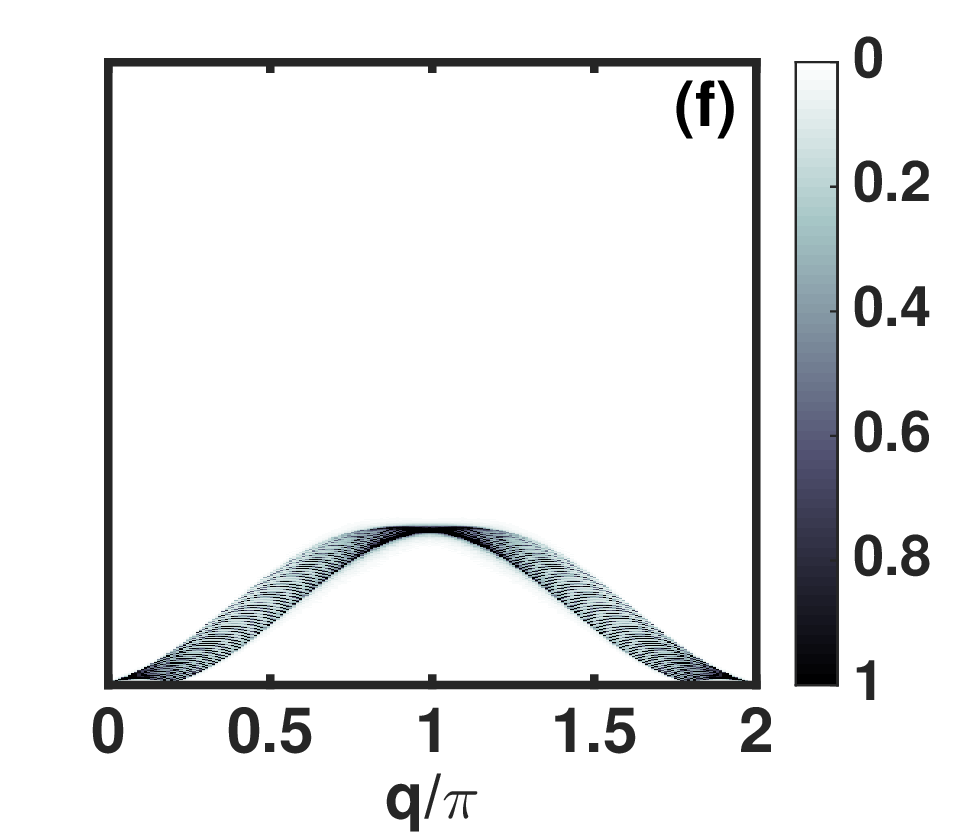,clip=0.4,
width=0.33\linewidth,height=0.32\linewidth,angle=0}
\caption{The intensity plots for the transverse DSFs $S^{-+}(q,\omega)$
from ($a_1$) to ($c_1$), $S^{+-} (q,\omega)$ from ($a_2$) to ($c_2$),
and for the longitudinal DSF $S^{zz}$ from $(d)$ to $(f)$
in the $q$-$\omega$ plane all with the same intensity scale.
$2m$ equals 0.2 in ($a_{1,2}$), 0.5 in ($b_{1,2}$), and 0.8 in  ($c_{1,2}$).
The $\delta$-function in Eq. \ref{eq:lehmann} is broadened via
a Lorenzian function
$\frac{1}{\pi} \gamma/[(\omega-E_\mu+E_G)^2+\gamma^2]$
with $\gamma=1/400$.
}
\label{fig:DSF}
\end{figure}

The intensity plots of the transverse DSSFs are presented in the $q$-$\omega$
plane in Fig. \ref{fig:DSF} at representative values of $2m$.
The spectra of $S^{-+}(q,\omega)$ exhibit the reminiscence of the
DCP modes at zero field \cite{DesCloizeaux1962}
shown in Fig.~\ref{fig:DSF} $(a_1)$, $(b_1)$, and $(c_1)$,
but are significant only
in the momentum interval of $2m\pi<q<2\pi-2m\pi$.
This can be understood intuitively in terms of the 1D Hubbard chain at
half-filling.
Although a weak coupling picture is employed below, charge gap already
opens at infinitesimal $U>0$ and there is no phase transition.
The gapless excitations are insensitive to the high energy charge
sector, hence, we expect the analysis below should also apply
to the case of AFM spin chains.
At magnetization $m$, the Fermi points for two spin components split
exhibiting the Fermi wavevectors
$k_{f_{\uparrow,\downarrow}}=\pi(\frac{1}{2}\pm m)$.
The minimum momentum for flipping a spin down to up is the difference
between $k_{f_{\uparrow,\downarrow}}$, i.e., $\Delta k_f=2m\pi$ or equivalently
$(1-m)2\pi$, and the energy cost is zero.
At small polarizations, $S^{-+}(q,\omega)$ is very coherent near $q=\Delta k_f$,
while as $q$ approaches $\pi$, it becomes a continuum.
The lower boundary of the continuum touches zero at $q=\pi$ corresponding to
flipping a spin-down at one Fermi point and adding it to the spin-up Fermi
point on the opposite direction.
The momentum interval for $S^{-+}$ shrinks as increasing polarization
and vanishes at the full polarization.

\begin{figure}
\centering\epsfig{file=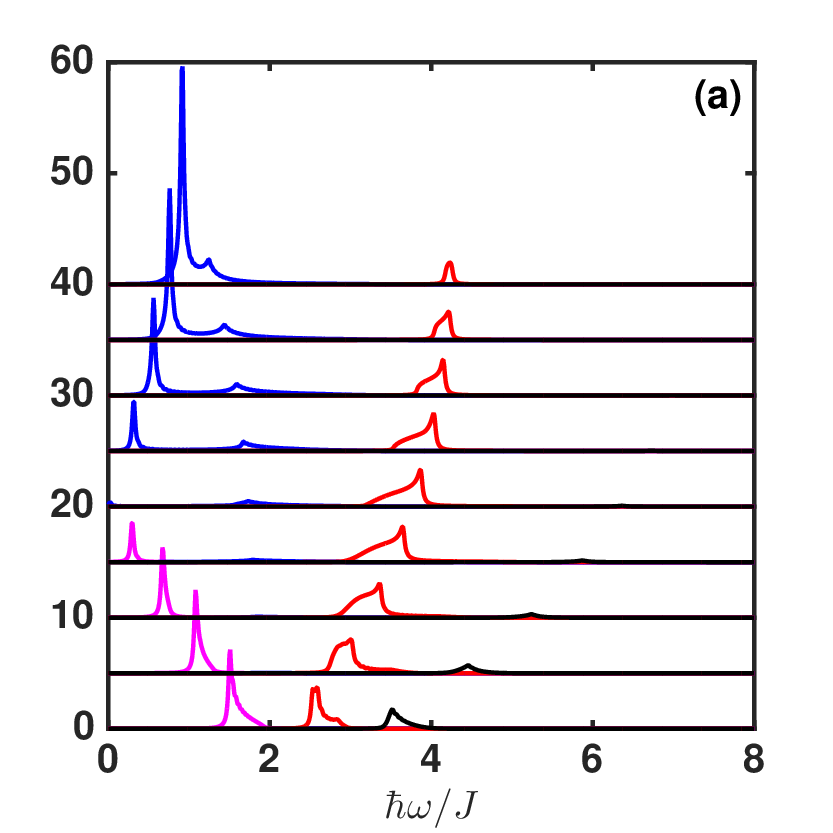,clip=0.9,
width=0.45\linewidth,angle=0}
\centering\epsfig{file=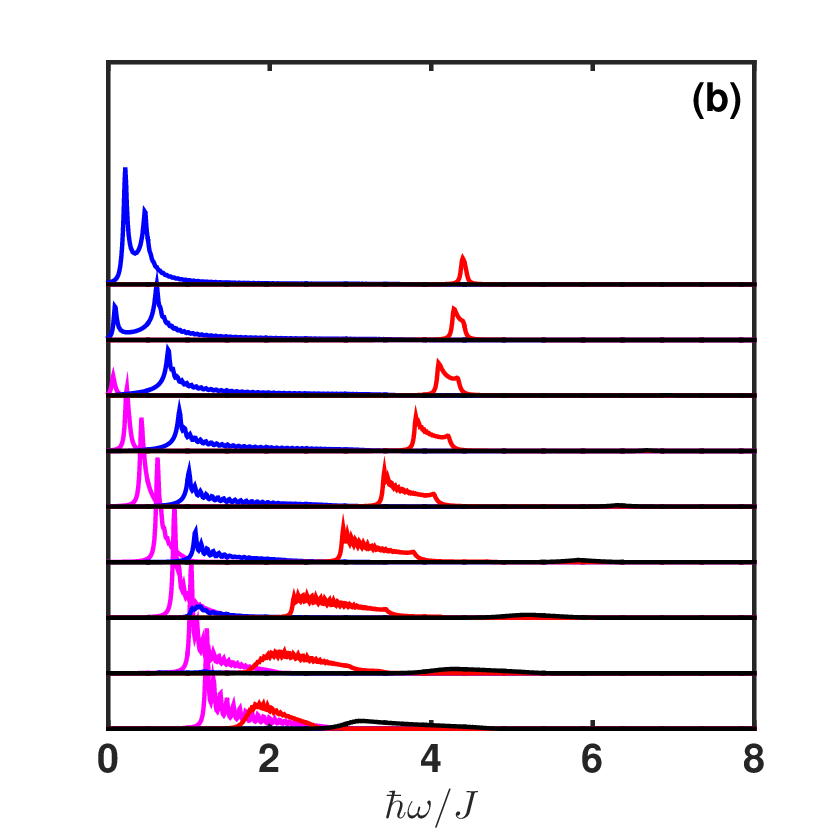,clip=0.9,
width=0.45\linewidth,angle=0}
\caption{Spectrum intensity evolution of $S^\perp(q,\omega)=S^{+-}(q,\omega)+
S^{-+}(q,\omega)$ $v.s.$ $\hbar \omega/J$ at ($a$) $q=\frac{\pi}{2}$, and
($b$) $q=\frac{3\pi}{4}$.
In ($a$) and ($b$), lines from bottom to up correspond to $2m$ varying
from $0.1$ to $0.9$ with the step of 0.1.
Contributions from psinon excitations in the $S^{-+}$ channel
are plotted in pink.
Psinon-antipsinon, 2-string and 3-string states in the $S^{+-}$  channel
are plotted in blue, red and black colors, respectively.
The broadening parameter $\gamma=1/50$.
}
\label{fig:intensity}
\end{figure}

The spectra of $S^{+-}(q,\omega)$ are presented in
Fig.~\ref{fig:DSF} ($a_2$), ($b_2$), and $(c_2$).
At small polarizations, the spectra resemble the DCP modes and further
split into three sectors.
Recall the ground state evolution as increasing polarization:
At $\Delta>1$, the ground state exhibits the N\'eel ordering at $m=0$, or,
the commensurate charge-density-wave (CDW) of particles.
With hole-doping, the ground state quantum-mechanically melts and
becomes incommensurate.
The low energy excitations are thus gapless, however, the intermediate
and high energy excitations still sense the gapped N\'eel state.
Applying $S^{-}(q)$ on $|G\rangle$ corresponds to adding back one particle.
A prominent spectra feature at low energy is the coherent Larmor precession
mode.
At $q=0$ and the isotropic case, the Larmor precession mode describes
the rigid body rotation with the eigenfrequency $\omega=h$
unrenormalized by interaction.
With anisotropy and away from $q=0$, it is renormalized by interaction
but remains sharp.
The antiferromagnetic coupling causes the downturn of the dispersion
touching zero at $q=\pm 2\pi m$, and then disappears.
The spectra around $q=\pi$ is incoherent as a reminiscence of the
two-spinon continuum in the zero-field DCP mode.
The intermediate and high energy spectra arise from the 2- and 3-string
states describing 2- and 3-particle bound states, respectively.
The energy separations among these three sectors are the reminiscence of
the spin gap of the N\'eel state.
As increasing polarization, the Larmor mode evolves to the magnon mode.
The states containing a pair of bounded magnons contribute to the upper
dynamical branch, which are high energy modes since the coupling
is anti-ferromagnetic.

We explicitly display the transverse DSF intensities {\it v.s.} $\hbar \omega/J$
from small to large polarizations at two representative wavevectors
$q=\frac{\pi}{2}$ and $\frac{3}{4}\pi$ shown in Fig. \ref{fig:intensity}.
The peaks reflect the large-weight region of the spectra in Fig. \ref{fig:DSF}.
The low frequency peaks are typically from the 2-particle excitations
of the $1\psi\psi$ and $1\psi\psi^*$ states.
In contrast, the intermediate and high frequency peaks are based on
multi-particle string state excitations.
For example, the 2-string states $1\chi^{(2)}1\psi\psi$ are 4-particle
excitations composed of a 2-particle bound state and a psinon-psinon
pair excitations.
Therefore, the string-state-based peaks are typically more smeared
than the low frequency peaks.

\begin{figure}[t]
\begin{center}
\includegraphics[width=0.6\linewidth]{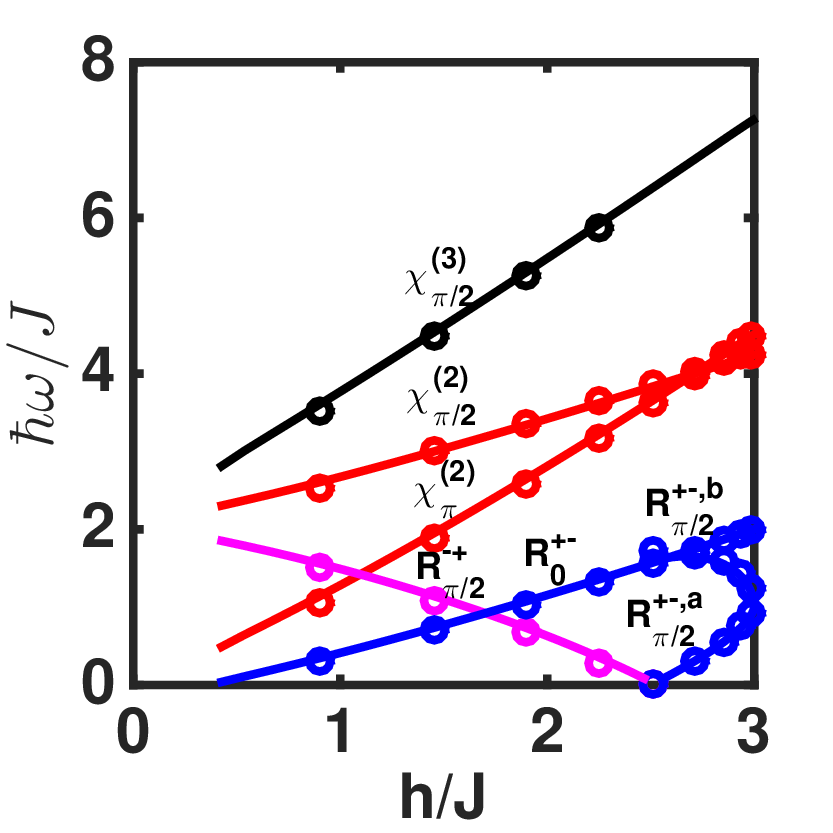}
\end{center}
\caption{
The evolution of peaks in DSSFs of $S^{+-}$ and $S^{-+}$ at different
momenta versus magnetic field $h$ with lines of peaks marked by $\chi^{(3)}_{\pi/2}$, $\chi^{(2)}_{\pi/2}$, $R_{\pi/2}^{-+}$, $\chi_\pi^{(2)}$, $R^{+-}_{0}$, $R^{+-,a}_{\pi/2}$, and
$R^{+-,b}_{\pi/2}$.
The pink, blue, red and black colors correspond to real states in $S^{-+}$,
real states in $S^{+-}$, two-string states in $S^{+-}$ and three-string
states in $S^{+-}$, respectively.
The hollow circles represent the peak positions extracted from DSSF
spectral figures similar to Fig. \ref{fig:intensity}.
The solid lines are determined by solving the energies of the Bethe eigenstates with the largest weight values around the spectral peaks.
}
\label{fig:bright_line}
\end{figure}

The evolutions of the spectral peaks at momenta $0$, $\frac{\pi}{2}$ and $\pi$
as tuning the magnetic field are displayed in Fig. \ref{fig:bright_line}.
We identify the lines of peaks
\bea
\chi^{(3)}_{\pi/2}, ~ \chi^{(2)}_{\pi/2}, ~ R^{-+}_{\pi/2}, ~
\chi^{(2)}_{\pi}, ~ R^{+-}_{0}, ~ R^{+-,a}_{\pi/2}, ~ R^{+-,b}_{\pi/2},
\eea
where the subscripts denote the corresponding momenta,
and $a, b$ label the two branches of peaks in $R^{+-}_{\pi/2}$.
The positions of the hollow circles are determined as follows:
We locate the spectral peak frequency position of each channel at the
corresponding momenta.
Further, the Bethe states with the largest spectral weight and the
associated quantum numbers can be identified, and the corresponding
eigen-energies are plotted by solid lines in Fig. \ref{fig:bright_line}
which indeed pass through the hollow circles.

Here we briefly summarize these states, with details included in
Appendix \ref{sect:state_peak}.
For the 3-string states $\chi^{(3)}_{\pi/2}\psi\psi^*$, which
consist a 3-string, one psinon, and one anti-spinon, the Bethe
eigenstate at the peak position of $S^{+-}(q,\omega)$ is characterized
with the partition of momenta as
\bea
k_{\chi^{(3)}}=\pi(1-m),~k_{\psi}=0, ~k_{\psi^*}=\pi (\frac{1}{2}+m),
\eea
where $k$ denotes the momentum, $m$ is the magnetization per site,
and the subscripts in $k$ represents the type of the excitation.
For the 2-string states $\chi^{(2)}_{\pi/2}\psi\psi^*$, the momentum
partition is
\bea
k_{\chi^{(2)}}=\pi(1+m), ~~ k_{\psi}=0, ~~ k_{\psi^*}=\pi(\frac{3}{2}-m).
\eea
Similarly, that of $\chi^{(2)}_\pi$ is
\bea
k_{\chi^{(2)}}=\pi(1-2m), ~~k_{\psi}=k_{\psi^*}=\pi(\frac{1}{2}+m).
\eea
The spectral peaks from states of real momenta are located at boundaries
of the two-particle continuum, which is an analogue of the X-ray edge singularity\cite{Pustilnik2006,Pereira2008}.
The following excitations, their momentum partitions are
\bea
&& R^{-+}_{\pi/2}: k_{\psi_1}=\pi(\frac{1}{2}+m), ~ k_{\psi_2}=\pi(1-m)
 \nonumber \\
&& R^{+-}_{0}: k_{\psi}=\pi(\frac{1}{2}+m), ~ k_{\psi^*}=\pi(\frac{1}{2}-m)
\nonumber \\
&& R^{+-,a}_{\pi/2}:
k_{\psi}=\pi(\frac{1}{2}+m), ~ k_{\psi^*}=\pi(1-m) \nonumber \\
&& R^{+-,b}_{\pi/2}:
k_{\psi}=\pi(\frac{3}{2}-m), ~k_{\psi^*}=\pi m
\eea
In all of above cases, to obtain the momentum transfer $q$ in Eq. (\ref{eq:lehmann}),
an additional $\pi$ shift must be added since $S^{+-}$ and
$S^{-+}$ change the  ground state magnetization by $1$.
It is interesting to note that several lines in Fig. \ref{fig:bright_line} exhibit nearly linear relation.
The identification of the above Bethe eigenstates is useful for an analytic analysis of the spectral peaks in the thermodynamic limit,
which will be left for a more careful future study.

\subsection{More discussions on transverse DSFs}
\label{sect:more}

\begin{figure}
\centering\epsfig{file=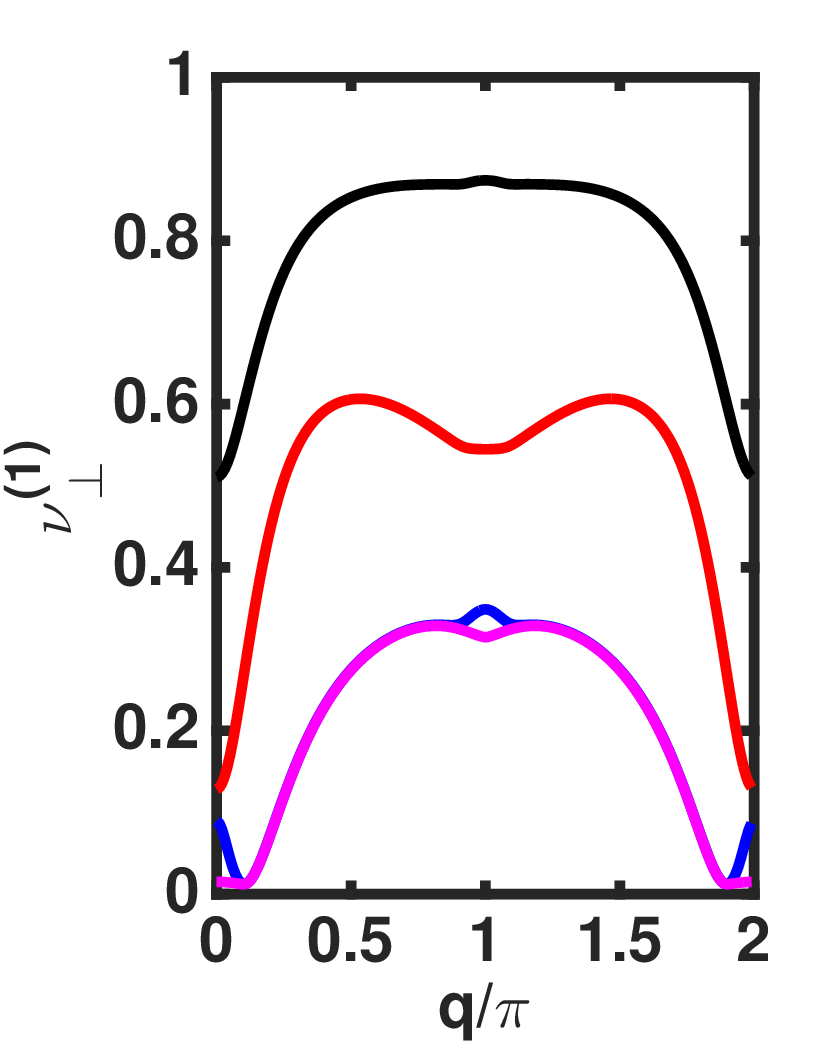,clip=1.5,width=0.4\linewidth,angle=0}
\caption{The momentum-resolved FFM ratios at $2m=0.1$.
The pink, blue, red and black curves represent cumulative results
by including the psinon states $n\psi\psi$ ($n=1,2$) in $S^{-+}$,
the psinon-antipsinon states $n\psi\psi^*$ ($n=1,2$), the
2-string states and 3-string states in $S^{+-}$, respectively, as before.
The anisotropy $\Delta=2$, and system size $N=200$.
}
\label{fig:first_200_10}
\end{figure}

\begin{figure}
\centering\epsfig{file=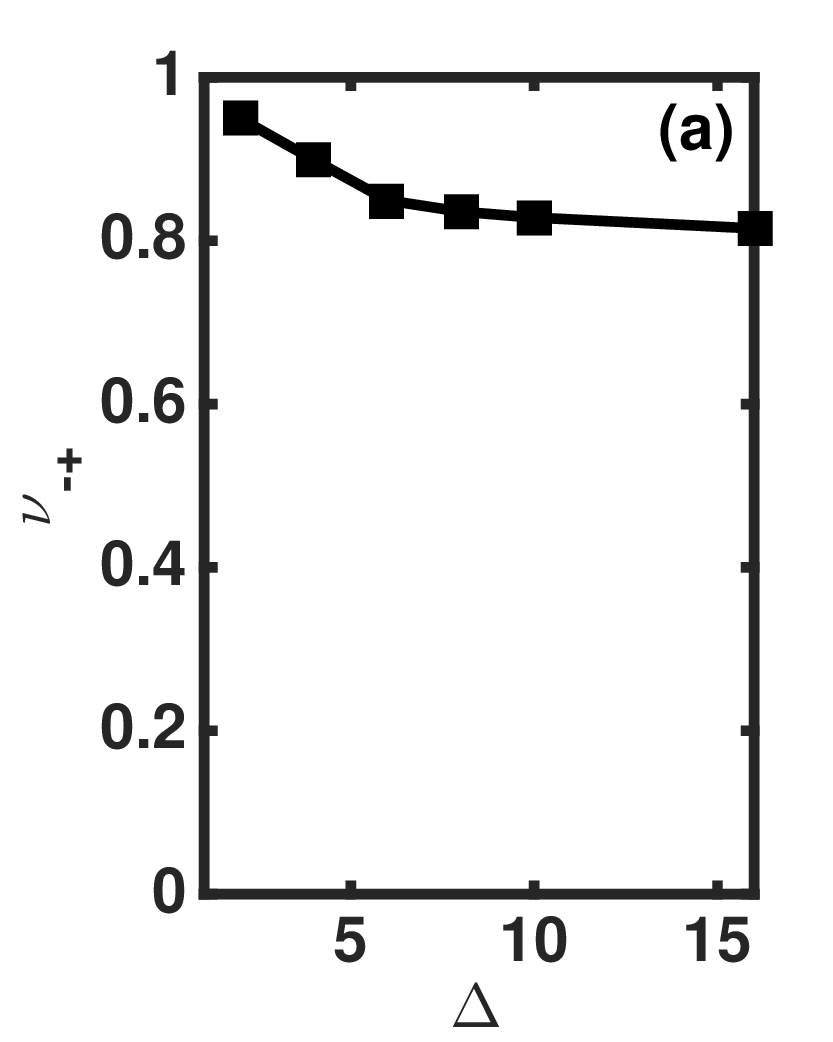,clip=0.6,
width=0.4\linewidth,angle=0}
\centering\epsfig{file=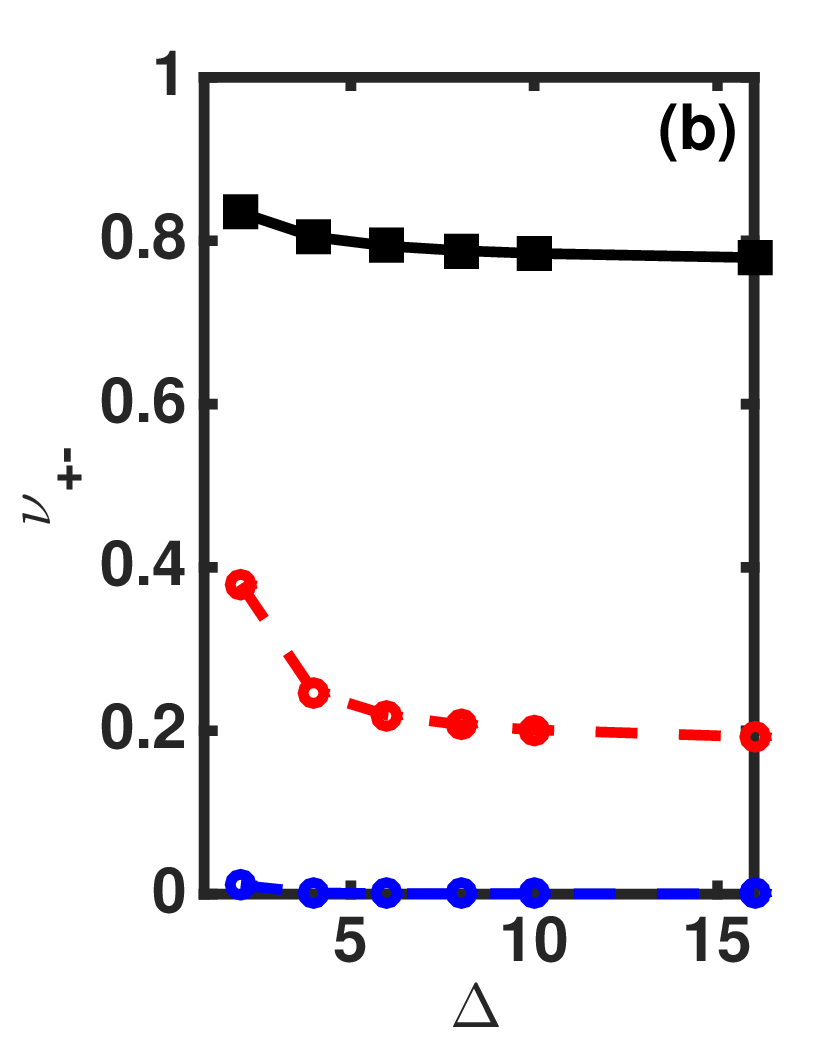,clip=0.6,
width=0.4\linewidth,angle=0}
\caption{The $\Delta$-dependence of the ratios of momentum integrated
intensity (a) $\nu_{-+}$, and (b) $\nu_{+-}$.
The parameter values are $N=200$ and $2m=0.05$.
In (a), the contributions from $1\psi\psi$ and $2\psi\psi$ states are included.
In (b), the blue, red, and black curves display the results
by cumulatively including the psinon-antipsinon, 2-string and 3-string
contributions in $S^{+-}$, respectively.
}
\label{fig:delta}
\end{figure}

To further investigate the behavior of the transverse DSFs near
the critical piont,
we present the FFM ratio at $2m=0.1$ in Fig. \ref{fig:first_200_10}.
A high saturation level ($>80\%$) is reached for most momenta,
however, near $q=0$, $\nu^{(1)}_{\perp}(q)$ drops to about $50\%$.
This indicates that there may exist unknown modes with significant
weights around zero momentum.


We also investigate the relation of the transverse DSFs with the
anisotropy parameter $\Delta$ as shown in Fig. \ref{fig:delta}.
We use the momentum-integrated sum rule \cite{Hohenberg1974} is
\bea
R_{a\bar a}= \frac{1}{N} \sum_{q}
\int_0^{\infty} \frac{d\omega}{2\pi} S^{a,\bar a} (q,\omega)
= \frac{1}{4} + \frac{m}{2} c_a,
\eea
where $c_a=\pm 1, 0$ for $a=\pm$ and $z$, respectively.
The saturation ratio for the integrated intensity is defined as
$\nu_{a\bar{a}}=\tilde{R}_{a\bar a}/R_{a\bar a}$ with
$a=\pm$ and $z$, where $\tilde{R}_{a\bar a}$ is from the
partial summations  over the selected excitations.

The small polarization regime is considered for the example of
$2m=0.05$, and the anisotropy parameter $\Delta$ takes values
of $2, 4,  6, 8, 10,$ and $16$.
For $S^{-+}$, the contributions to $\nu_{-+}$ from the $1\psi\psi$ and
$2\psi\psi$ states drop to about $80\%$ as increasing $\Delta$,
and the absent weights may arise from string states.
For $S^{+-}$, the dominance of three-string states continuously enhances
as increasing $\Delta$ towards the Ising limit.
While the three-string states become increasingly dominant as approaching
the critical line, it is known that there are no strings of length longer
than two in the zero magnetic field case \cite{Babelon1983,Woynarovich1982}.
A more careful investigation to the regime of very small magnetization
will be deferred to a future work.

\section{The longitudinal dynamic spin structure factor}
\label{sect:longitudinal}
In this section, we continue to present the longitudinal
DSSF, i.e., $S^{zz}(q,\omega)$ of Eq.~(\ref{eq:Hamiltonian}),
and also check the saturation level by using sum rules.

\subsection{The momentum--resolved ratios of the longitudinal DSSF}
\label{sect:saturation_Szz}

\begin{figure}
\centering\epsfig{file=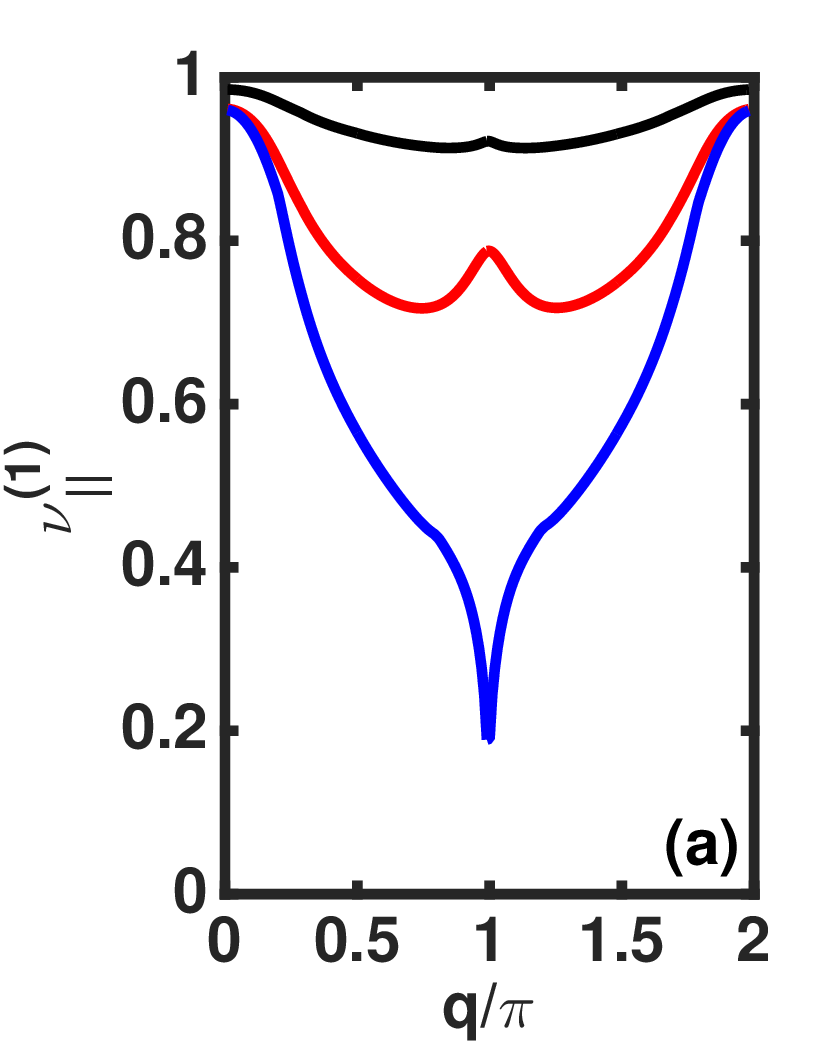,clip=1.5,
width=0.35\linewidth,height=0.4\linewidth,angle=0}
\centering\epsfig{file=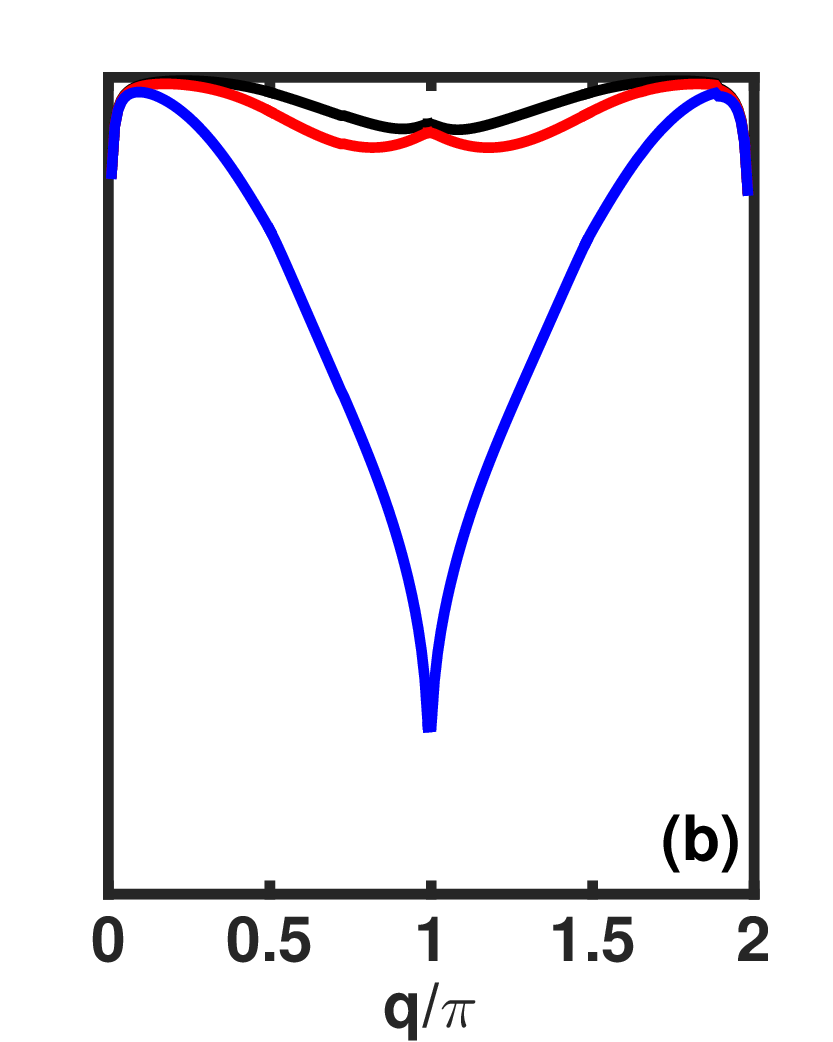,clip=1.5,
width=0.30\linewidth,height=0.4\linewidth,angle=0}
\centering\epsfig{file=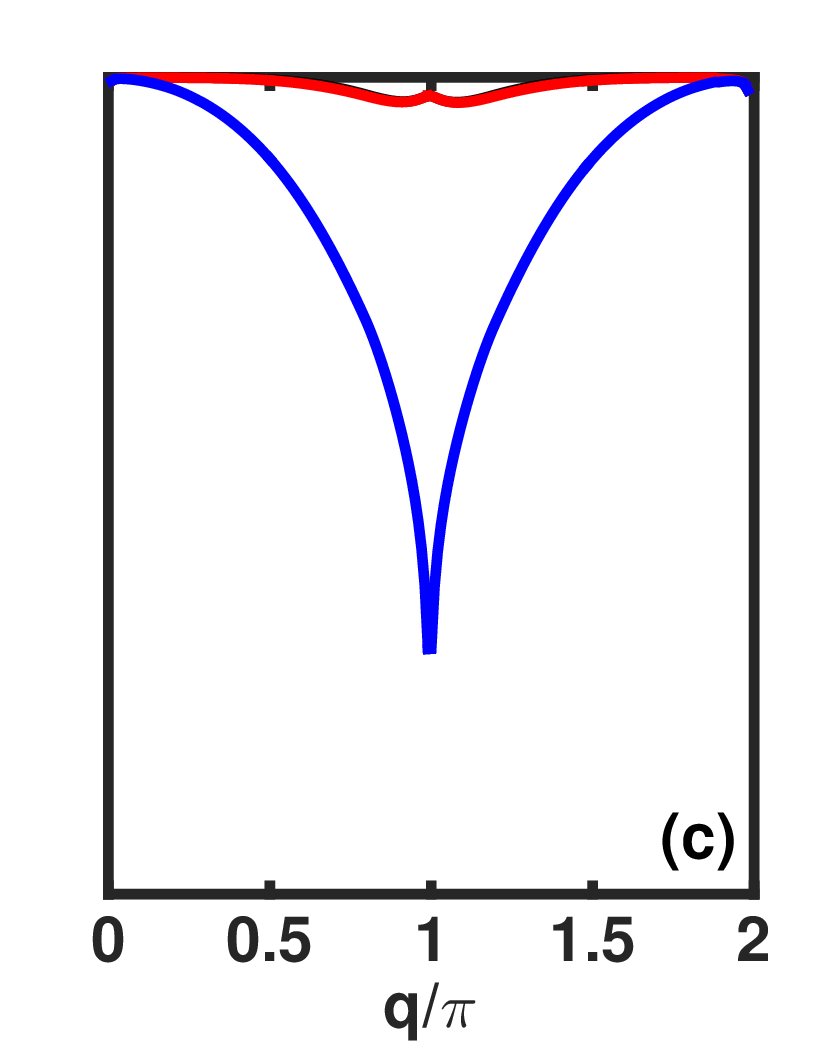,clip=1.5,
width=0.30\linewidth,height=0.4\linewidth,angle=0}
\caption{The momentum-resolved FFM $\nu^{(1)}_{||} (q)$ ratios from
$(a)$ to $(c)$, and the intensity plots from $(d)$ to $(f)$ for the
longitudinal DSF $S^{zz}$.
$2m$ equals $0.2$ in $(a)$ and $(d)$, $0.5$ in $(b)$ and $(e)$,
and $0.8$ in $(c)$ and $(f)$, respectively.
In $(a)$, $(b)$ and $(c)$, the blue, red and black lines are
cumulative results by including $1\psi\psi^{*}$, $2\psi\psi^{*}$, and
$1\chi^{(2)}1\psi\psi$ excitations.
The broadening parameter in the intensity plots is $\gamma=1/400$.
}
\label{fig:szz}
\end{figure}

The momentum resolved longitudinal first frequency moment (FFM)
sum rule is known as
\bea
W_\pp(q) =
\int_{0}^\infty \frac{d \omega}{2\pi} \omega S^{zz} (q,\omega )
=(1-\cos q) \alpha_\pp \cite{Mossel2008},
\eea
where $\alpha_\pp=-e_0+\Delta {\partial e_0}/{\partial \Delta}$.
We define the ratio of
$\nu^{(1)}_{\pp}(q)=\tilde{W}_{\pp}(q)/W_{\pp}(q)$ in the longitudinal channel, where again $\tilde{W}_{\pp}(q)$ is calculated
from the partial summations over the selected excitations.

The momentum-resolved ratios $\nu^{(1)}_{zz}(q)$ at representative polarizations and the intensities of $S^{zz}(q,\omega)$ are plotted in Fig.~\ref{fig:szz} after taking into account excitations of $1\psi\psi^{*}$, $2\psi\psi^{*}$,
and $1\chi^{(2)}1\psi\psi$ states.
Satisfactory saturation levels are obtained.

\subsection{The spectral weights}
The calculated spectra weights are plotted in Fig. \ref{fig:DSF} $d$, $e$,
and $f$ for $2m=0.2,0.5$ and $0.8$, respectively.
This quantity is equivalent to the dynamic density-density correlations
of a $1D$ interacting spinless fermion system through the Jordan-Wigner
transformation with the identification of
the Fermi wavevector $k_f=\frac{\pi}{2}(1-2m)$.

At small polarizations, the contribution of string states
dominates the high energy spectra branch.
The low energy excitations in the long wavelength regime are very
coherent due to the structure of 1D phase space, while those at $2k_f$
are incoherent, both of which can be described by the 1D Luttinger
liquid theory \cite{Voit1993}.
The high energy excitations are the reminiscence of the gapped excitonic
excitations in the commensurate N\'eel background.
As increasing polarization, particle filling touches the band bottom
where the band curvature is important, and thus the low energy coherent
excitations are suppressed and particle-hole continuum becomes more prominent.
When the ground state evolves further away towards the fully polarization,
the low energy excitations are more incoherent, and
the spectra from the string state excitations diminish.


\section{Discussion and Conclusion}
\label{sect:discussion}

We discussion the implication of our results for experiments.
The quasi-1D $\text{SrCo}_{2}\text{V}_{2}\text{O}_{8}$
AFM chain can be effectively described
by the XXZ model with parameters $\Delta =2, J = 3.55 \rm{meV}$,
and the Land\'{e} factor $g_z = 6.2$, and
the critical value of magnetic field is about $h_c=4T$
\cite{Wang2015a, Wang2018}.
The Brillouin zone of the material is folded into a fourth
due to its four-fold screw periodic structure,
hence the electronic spin resonance (ESR) measurements can detect the DSF of
$S^{+-} + S^{-+}$ at momenta
$0$, $\frac{\pi}{2}$, $\pi$ and $\frac{3\pi}{2}$,
in which $\frac{\pi}{2}$ and $\frac{3\pi}{2}$ are equivalent
due to the inversion symmetry.
Indeed, the ESR experiment on the
material $\text{SrCo}_{2}\text{V}_{2}\text{O}_{8}$ \cite{Wang2018} not only
confirms the real excitations but also for the first time clearly observes the
string excitations, in which the experimental results agree well with
our theoretical predictions in Fig.~\ref{fig:bright_line},
demonstrating a rare success of the strong-correlation
description for the real material from low to high energy region \cite{Wang2018}.
Furthermore, the quantity $1/2(S^{+-}+S^{-+})+S^{zz}$ can be compared with
inelastic neutron scattering experiments for the whole range of $(q,\omega)$.

Besides the spin system, the 1D bosonic system in the hard-core regime is equivalent to the spin-$\frac{1}{2}$ chain, which has been realized in cold
atom experiments \cite{Fukuhara2013Q}, and quantum dynamics of
two-magnon bound states has been measured \cite{Fukuhara2013}.
Our DSSF calculations and various identified excitations provide
helpful guidance to the experimental study of quantum spin dynamics
in these systems.


Although the above concrete calculations are based on the integrity of
the 1D spin-$\frac{1}{2}$ XXZ model, we believe that the underlying
physics at high energies is universal not limited to integrable
models.
Based on Fig. \ref{fig:melting} ($e$) and ($f$), we have explained the
physical picture of 2 and 3-string states, and the absence of 4-string
states.
Similar physics is also speculated in non-integrable models, such as
in the two-dimensional AFM XXZ model.
Under similar physical parameter set-ups, we would expect it is possible
to observe contributions from 2, 3, 4, and up to 5-magnon
clustering states, since in a two-dimensional geometry the coordination
number is 4.
Certainly for the 2D case, the method of Bethe ansatz will not be
possible, and the theory study will be deferred to a future
publication.

In summary, the zero temperature spin dynamics is studied for the
spin-$\frac{1}{2}$ AFM XXZ model in the longitudinal magnetic field.
We find that different dynamic branches are energetically separated,
which originate from various classes of excitations including
psinon-psinon and psinon-antipsinon pairs at low energy,
and string excitations at intermediate and high energies.
In particular, for $S^{+-}(q,\omega)$ at small magnetizations,
states with real rapidities contribute negligibly small
to the sum rule, and the 3-string states become more and more dominant
as approaching the critical line or increasing anisotropy.
These high-frequency spin dynamic features cannot be captured
within the low energy effective theory of the Luttinger liquid.
Our calculations provide important guidance for analyzing the 1D
spin dynamics experiments in both condensed matter and ultra-cold
atom systems.

\acknowledgments
We thank useful discussions with Matthew Foster.
W. Y., J. W., S. X. and C. W. are supported by the AFOSR
FA9550-14-1-0168.

\appendix

\section{Bethe ansatz in the axial regime}
\label{sect:BA}
In this section, we present the Bethe ansatz equations (BAE) and the Bethe
quantum number (BQN) structure.
We focus on the anti-ferromagnetic XXZ spin chain (Eq.~(1) in main text)
in the axial regime with $\Delta=\cosh \eta >1$.

In the method of the algebraic Bethe ansatz \cite{Faddeev1979}, the monodromy
matrix is a $2 \times 2$ matrix.
Its matrix entries $A(\lambda), B(\lambda), C(\lambda), D(\lambda)$ are
operators acting in the many-body Hilbert space of the spin chain.
By the virtue of the Yang-Baxter equation, all the transfer matrices
$T(\lambda)=A(\lambda)+D(\lambda)$ with different spectral parameter
$\lambda$'s commute, hence they can be simultaneously diagonalized.
The XXZ Hamiltonian can be expressed in terms of these transfer
matrices, and thus it shares common eigenstates with all the
transfer matrices.

A Bethe eigenstate with $M$ down-spins can be expressed as the result of
successively applying the magnon creation operators $B(\lambda_{j})$
($1 \leq j \leq M$) onto the reference state $\ket{F}=\otimes_{j=1}^{N}
\ket{\uparrow}_{j}$, as $\Pi_{j=1}^{M} B(\lambda_{j}) \ket{F}$.
The rapidities $\{ \lambda_{j} \}_{1\leq j\leq M}$ satisfy the Bethe ansatz
equations,
\bea
N \theta_{1}(\lambda_{j})=2 \pi I_{j} + \sum_{k=1}^{M} \theta_{2}
(\lambda_{j}-\lambda_{k}),
\label{eq:bethe ansatz eq}
\eea
where
\bea
\theta_{n}(\lambda)=2 \arctan (\frac{\tan(\lambda)}{\tanh(n\eta/2)})+
2\pi \floor{\frac{\Re(\lambda)}{\pi}+\frac{1}{2}}. \ \ \
\label{eq:theta}
\eea
The symbol $\floor{x}$ represents the floor function, which yields
the largerst integer less than or equal to $x$.

The rapidities can be either real or complex in general.
If all $\lambda_{j}$'s are real, then the corresponding state is called a
real Bethe eigenstate.
If there exist complex-valued $\lambda_{j}$'s, then the state is called
a string state\cite{Takahashi2005}, whose name comes from the pattern of
$\lambda_{j}$'s in the complex plane in the thermodynamic limit.
We will give a brief description in Appendix ~\ref{sect:string}.

For a chain with even number of sites, the ascending array of Bethe
quantum numbers $\{ I_{j} \}_{1 \leq j \leq M}$ take integer values
when $M$ is odd, and half-integer values when $M$ is even.
The total momentum of this state is
\bea
P=\pi M-\frac{2\pi}{N} \sum_{j=1}^{M} I_{j},
\eea
and the energy is
\bea
E=\sum_{j=1}^{M} \frac{\sinh^{2}(\eta)}{\cosh \eta -\cos(2 \lambda_{j})}.
\eea

In the subspace with a fixed value of $S_{T}^{z}$, there exist
$M=\frac{N}{2}-S^{z}_{T}$ down-spins.
In this sector, the BQN of the lowest energy state
are given by
\bea
I_{j}=-\frac{M+1}{2}+j,\,\,\, 1 \leq j \leq M.
\label{eq:ground Bethe num}
\eea


As for the excited states, the BQN can be grouped into
certain patterns by examining how they can be obtained through
modifying those in the ground state given in Eq. (\ref{eq:ground Bethe num}).
We consider two different classes of excited states with purely real
rapidities.
Eigenstates with $n$-pair of psinons are denoted $n\psi\psi$
\cite{Karbach2000}, and their
Bethe quantum numbers $\{I_{j}\}_{1\leq j\leq M}$ satisfy
\bea
-\frac{M-1}{2}-n \leq I_{j} \leq \frac{M-1}{2}+n,
\eea
where either $I_{1}=-\frac{M-1}{2}-n$ or $I_{M}= \frac{M-1}{2}+n$
to avoid over-counting.
Another class of solutions are called $n$-pair of psinon-anti-psinon
states denoted $n\psi\psi^{*}$.
Among their $M$ Bethe quantum numbers $I_j$'s, $M-n$ of them
lying within the range $[-\frac{M-1}{2},\frac{M-1}{2}]$, and the
remaining $n$ ones lying outside
\cite{Karbach2000}.

\section{The Bethe-Gaudin-Takahashi equations for string states}
\label{sect:string}
The rapidities of the BAE can take complex values, and
the corresponding solutions are called string states\cite{Takahashi2005}.
The string ansatz assumes that the complex rapidities form the string
pattern described below.

For a single $n$-string of complex rapidities,
\bea
\lambda^{n}_{j}=\lambda^{(n)}+i (n+1-2j)\frac{\eta}{2}, ~~~
1\le j \le n,
\label{eq:string rapidity}
\eea
where $\lambda^{(n)}$ and $\eta$ are real numbers, and $j$
is the rapidity index inside the string.
For a finite system the distribution of rapidities does not exactly follow
Eq. (\ref{eq:string rapidity}).
The deviations become exponentially suppressed as enlarging system size,
and the string ansatz is asymptotically exact in the thermodynamic limit.
Then a general Bethe eigenstate with $M$ rapidites is a collection
of $M_{n}$ $n$-strings, where $\sum_{n} n M_{n}=M$.
A real Bethe eigenstate can be also viewed as a collection of $M$
$1$-strings in this language.

The BAE Eq. (\ref{eq:bethe ansatz eq}) becomes singular in thermodynamic limit for a string state with the rapidity pattern of
Eq. (\ref{eq:string rapidity}).
Their regularized version is called the Bethe-Gaudin-Takahashi (BGT)
equations \cite{Takahashi2005}, which only contain the common real
part $\lambda^{(n)}$
\bea
N \theta_{n} (\lambda_{\alpha}) = 2\pi I^{(n)}_{\alpha} +\sum_{(m,\beta) \neq (n,\alpha)} \Theta_{nm} (\lambda^{(n)}_{\alpha}-\lambda^{(m)}_{\beta}), \ \ \,
\eea
with  $1\leq \alpha \leq M_{n}, \,\,\, 1 \leq \beta \leq M_{m}$,
where
\bea
\Theta_{nm}&=&(1-\delta_{nm})\theta_{|n-m|}+2 \theta_{|n-m|+2}+...\nn\\
&+&2 \theta_{n+m-2} +\theta_{n+m},
\eea
and $\theta_{n}$ is defined in Eq. (\ref{eq:theta}).
The momentum of such a state is
\bea
P=\pi \sum_n M_n-\frac{2\pi}{N} \sum_{n\alpha} I^{(n)}_{\alpha}
\label{eq:string_momentum}
\eea
and the energy is
\bea
E=\sum_{n\alpha}\frac{\sinh(\eta) \sinh(n\eta)}{\cosh( n \eta)-\cos(2\lambda^{(n)}_{\alpha})}.
\eea

The general rules for determining BQN for distinct eigenstates are
rather complicated \cite{Mossel2010}.
Since only Bethe eigenstates with up to only two
types of strings are considered in this article, we only
present the rules for these special cases below \cite{Mossel2010}.

Consider a string state with $M_{m}$ $m$-strings and $M_{n}$ $n$-strings,
where $M=m M_{m}+n M_{n}$.
Without loss of generality, we assume $m<n$.
The BQN for the $m$-strings are within the sets of
\be
A^{(m)}_{i}=\{-\frac{W_{m}-1}{2}+i \leq I^{m}_{j} \leq
\frac{W_{m}-1}{2}+i,1\leq j \leq M_{m} \},
\ee
where
\bea
W_{m}=N-2mM_{n}-(2m-1)M_{m},
\eea
and $0\leq i \leq 2m-1$.
For the $n$-strings, the BQN are within the sets of
\be
A^{(n)}_{i} = \{-\frac{W_{n}-1}{2}+i \leq I^{n}_{j} \leq \frac{W_{n}-1}{2}+i,1\leq j \leq M_{n}\},
\ee
where
\bea
W_{n}=N-2mM_{m}-(2n-1)M_{n},
\eea
and $0\leq i \leq 2n-1$.
Not all these BQN yield distinct Bethe eigenstates.
To remove equivalent sets of BQN giving same
eigenstates, we need to exclude those simultaneously satisfying
the following two conditions
\bea
I^{(m)}_{1}&\leq& -\frac{W_{m}-1}{2}+2m-1, \nn \\
I^{(n)}_{M_{n}} &\geq& \frac{W_{n}-1}{2} +2n-(2m-1).
\eea

In the following, the presence of the rules of Bethe quantum numbers
for $2$-string and $3$-string states are combined together to reduce
the content.
We list the rules for the BQN of the string states calculated in the main text.
In the following formulae, $n=2$ or $3$.
The rule for $1\chi^{(n)}1\psi\psi$ state is
\bea
&-\frac{N-2M}{2}\leq &I^{(n)} \leq \frac{N-2M}{2}+2n-1,\nn\\
&-\frac{M-n+1}{2}+i\leq &I^{(1)}_{j} \leq \frac{M-n+1}{2} +i,
~~ 1\leq j \leq M-n, \nn \\
\eea
in which $i$ is an integer.
The DSF intensity distribution must be symmetric with respect to
$π$ since the system possesses inversion symmetry.
It is possible for states with $i=0$ to be transformed
to those with $i\neq 0$ under inversion, which must also be included.

For the excitations of the type of $1\chi^{(n)}1\psi\psi^{(*)}$,
the rule for the $I^{(n)}$ part is the same, while that for real
rapidities is
\bea
&-\frac{M-n-1}{2}+i\leq& I^{(1)}_{j_{l}} \leq \frac{M-n-1}{2}+i,
\nn\\
&&\,\,\,\,\,\,\,\,\,\,\,\,1\leq l\leq M-n-1,\nn\\
&-\frac{N-M+n-3}{2}\leq &I^{(1)}_{j_{M-n}}\leq -\frac{M-n-1}{2}-1+i,\text{ or }\nn\\
& \frac{M-n-1}{2}+1+i\leq& I^{(1)}_{j_{M-n}}\leq \frac{N-M+n-3}{2}+1,
\eea
where $I^{(1)}_{j}$'s should be arranged in an ascending array,
and $-(2n-1)\le i\le 2n-1$ again for the purpose of symmetrization.
The BQN need to be excluded
if they simultaneously satisfy the following two conditions
$I^{(n)}\geq \frac{N-2M}{2}+2n-2$ and $I^{(1)}_{1}\leq -\frac{N-M+n-3}{2}+1$
to avoid overcounting as mentioned above.

\section{The determinant formulae}
\label{sect:determinant}

To carry out the DSF calculation, the normalized Bethe state and the
matrix element of spin operators are needed.
The normalized state of $\Pi_{j=1}^{M} B(\lambda_{j}) \ket{F}$ is denoted
as $\ket{\{ \lambda_{j} \}_{1\leq j \leq M}}$ below.
The matrix entries $\bra{\{ \mu_{k} \}_{1\leq k \leq M+1}}
S^{a}_{q} \ket{\{\lambda_{j}\}_{1\leq j \leq M}}$ can be formulated
into determinant forms \cite{Kitanine1999}, which greatly facilitates
both analytical and numerical calculations.

\begin{widetext}
\subsection{Real states in the axial regime}
We first present the determinant formulae for the real Bethe state.
Since $|\bra{\{ \mu_{k} \}_{1\leq k \leq M+1}}
S^{-}_{q} \ket{\{\lambda_{j}\}_{1\leq j \leq M}}|^2
=|\bra{ \{\lambda_{j}\}_{1\leq j \leq M}}
S^{+}_{-q} \ket{\{ \mu_{k} \}_{1\leq k \leq M+1}}|^2$,
we only present the matrix element for $S^-_q$ and $S^z_q$.

The transverse matrix element can be expressed as
\bea
|\bra{\{ \mu \}} S^{-}_{q} \ket{ \{ \lambda \}} |^{2}&=& N
\delta_{P(\{ \lambda \})-P(\{ \mu \}),q} |\sin{i\eta}|
\frac{\Pi_{k=1}^{M+1} |\sin(\mu_{k}-i\eta/2)|^{2} }{\Pi_{j=1}^{M} |\sin(\lambda_{j}-i\eta/2)|^{2}}
\nn \\
&\times& \frac{1}{\Pi_{k \neq k^{'}} |\sin(\mu_{k}-\mu_{k^{'}}+i\eta)
|\Pi_{j \neq j^{'}} |\sin(\lambda_{j}-\lambda_{j^{'}}+i\eta)| }
\frac{|\det H^{-}|^{2}}{| \det \Phi(\{ \mu \})  \det \Phi(\{ \lambda \})  |}.
\eea
in which $H^-$ is an $(M+1)\times (M+1)$ matrix.
For $1\leq k\leq M+1,\,\,\,1\leq j\leq M$,
\bea
H^{-}_{kj} &=& \frac{1}{\sin(\mu_{k}-\lambda_{j})}
[   \Pi_{l=1(l\neq k)}^{M+1} \sin(\mu_{l}-\lambda_{j}+i\eta)
- (\frac{\sin(\lambda_{j}-i \eta/2)}{\sin(\lambda_{j}+i \eta/2)})^{N}
\Pi_{l=1(l\neq k)}^{M+1} \sin(\mu_{l}-\lambda_{j}-i\eta)];
\eea
and for $1\leq k\leq M+1$,
\bea
H^{-}_{k,M+1}= \frac{1}{\sin(\mu_{k}+i\eta/2) \sin(\mu_{k}-i\eta/2)}.
\eea

For the longitudinal matrix element, the expression for
$\bra{\{ \mu_{k} \}_{1\leq k \leq M}}
S^{z}_{q} \ket{\{\lambda_{j}\}_{1\leq j \leq M}}$ is
\bea
|\bra{\{ \mu \}} S^{z}_{q} \ket{ \{ \lambda \}} |^{2}
&=& \frac{N}{4} \delta_{P(\{ \lambda \})-P(\{ \mu \}),q}  ~
\Pi_{k=1}^{M}  |\frac{ \sin(\mu_{k}-i\eta/2) }{\sin(\lambda_{j}-i\eta/2)}|^{2}
\nn \\
&\times&
\frac{1}{\Pi_{k \neq k^{'}} |\sin(\mu_{k}-\mu_{k^{'}}+i\eta)|  \Pi_{j \neq j^{'}}
|\sin(\lambda_{j}-\lambda_{j^{'}}+i\eta)| } ~
\frac{|\det (H-2P)|^{2}}{| \det \Phi(\{ \mu \})
\det \Phi(\{ \lambda \})  |},
\eea
in which the $M\times M$ matrices $H$ and $P$ are given by
\bea
H_{kj} &=& \frac{1}{\sin(\mu_{k}-\lambda_{j})}
[   \Pi_{l=1(l\neq k)}^{M} \sin(\mu_{l}-\lambda_{j}+i\eta)
- (\frac{\sin(\lambda_{j}-i \eta/2)}{\sin(\lambda_{j}
+i \eta/2)})^{N} \Pi_{l=1(l\neq k)}^{M} \sin(\mu_{l}-\lambda_{j}-i\eta)],
\eea
and
\be
P_{kj}=\frac{\Pi_{l=1}^{M} \sin(\lambda_l-\lambda_j-i\eta)}{\sin(\mu_{k}+i\eta/2)\sin(\mu_{k}-i\eta/2)},
~~\mbox{for}~~ 1\leq k\leq M,\,\,\,1\leq j\leq M.
\ee
The off-diagonal matrix elements $\Phi_{jk}$ at $(j\neq k)$
is
\bea
\Phi_{jk}&=&  \frac{\sin(2i\eta)}{\sin(\lambda_{j}-\lambda_{k}-i\eta)
\sin(\lambda_{j}-\lambda_{k}+i\eta)},
\label{eq:phi1}
\eea
and the diagonal matrix element $\Phi_{jj}$ is
\bea
\Phi_{jj}&=&N \frac{\sin(i\eta)}{\sin(\lambda_{j}-i\eta/2)
\sin(\lambda_{j}+i\eta/2)} -\sum_{l=1,l\neq j}^{M}
\frac{\sin(2i\eta)}{\sin(\lambda_{j}-\lambda_{l}-i\eta)
\sin(\lambda_{j}-\lambda_{l}+i\eta)}.
\label{eq:phi2}
\eea

\subsection{The reduced determinant formule for string states}
In calculating the DSFs, if we directly plug in the rapidities
of the string state solutions into Eqs.~(\ref{eq:phi1}, \ref{eq:phi2}),
the matrix $\Phi$ becomes singular.
The L'Hospital's rule must be applied to remove the singularities
\cite{Mossel2010}.
The reduced matrix $\Phi^{(r)}$ is defined by \cite{Mossel2010}
\bea
\Phi^{(r)}_{n\alpha,n\alpha}&=& N\sum_{j=1}^{n}
[\frac{\sin(i\eta)}{\sin(\lambda^{(n\alpha)}_{j}-i\eta/2)
\sin(\lambda^{(n\alpha)}_{j}+i\eta/2)}
-\sum_{k=1(k\neq n\alpha j, j\pm 1)}^{M}
\frac{\sin(2 i\eta)}{\sin(\lambda^{(n\alpha)}_{j}-\lambda_{k}-i\eta)
\sin(\lambda^{(n\alpha)}_{j}-\lambda_{k}+i\eta)}
\nn \\
&+&\sum_{l=1(l\neq j,j\pm1)}^{n} \frac{\sin(2 i\eta)}{\sin(\lambda^{(n\alpha)}_{j}-\lambda^{(n\alpha)}_{l}-i\eta)
\sin(\lambda^{(n\alpha)}_{j}-\lambda^{(n\alpha)}_{l}+i\eta)} ], \nn \\
\Phi^{(r)}_{n\alpha,m\beta}&=&
\sum_{j=1}^{n}\sum_{k=1}^{m} \frac{\sin(2i\eta)}{\sin(\lambda^{(n\alpha)}_{j}-\lambda^{(m\beta)}_{k}-i\eta) \sin(\lambda^{(n\alpha)}_{j}-\lambda^{(m\beta)}_{k}+i\eta)},\,\,\, n\alpha \neq m\beta,
\eea
in which $\lambda^{(n\alpha)}_{j}=\lambda^{(n\alpha)}+i(n+1-2j)\eta/2$,
where $\lambda^{(n\alpha)}$ is the common real part of the $\alpha$'th
length-$n$ string.

The formula for $|\bra{\{\mu\}}  S^{-}_{q} \ket{\{ \lambda \}}  |^{2}$,
where $\ket{\{ \mu \}}$ is a string state, $\ket{\{ \lambda \}}$
a real Bethe eigenstate, is given by
\bea
|\bra{\{\mu\}}  S^{-}_{q} \ket{\{ \lambda \}}  |^{2}&=& N
\delta_{P(\{ \lambda \})-P(\{ \mu \}),q}
\frac{|\sin(i\eta)|}{\Pi_{n} (|\sin^{n-1}(2i\eta)|)^{M_{n}}}
~\frac{\Pi_{k=1}^{M+1} |\sin(\mu_{k}+i\eta/2)| }{ \Pi_{j=1}^{M} |\sin(\lambda_{j}+i\eta/2)| }
~
\frac{1}{\Pi_{j\neq j^{'}} |\sin(\lambda_{j}-\lambda_{j^{'}}+i\eta ) |} \nn \\
&\times& \frac{1}{\Pi_{m\beta l \neq n\alpha l^{'},l^{'}\pm1}  |\sin(\mu^{(n\alpha)}_{l}-\mu^{(m\beta)}_{l^{'}}+i\eta)   |}
~
\frac{|\det H^{-}|^{2}}{|\det\Phi(\{\lambda\})|\cdot|\det \Phi^{r}(\{\mu\})|}.
\eea

The expression for $|\bra{\{\mu\}}  S^{z}_{q} \ket{\{ \lambda \}}  |^{2}$ can be obtained similarly, as
\bea
|\bra{\{\mu\}}  S^{z}_{q} \ket{\{ \lambda \}}  |^{2}&=& \frac{N}{4}
\delta_{P(\{ \lambda \})-P(\{ \mu \}),q}
\frac{1}{\Pi_{n} (|\sin^{n-1}(2i\eta)|)^{M_{n}}}
~\Pi_{j=1}^{M} |\frac{\sin(\mu_{j}+i\eta/2)}{\sin(\lambda_{j}+i\eta/2)} |^{2}
~
\frac{1}{\Pi_{j\neq j^{'}} |\sin(\lambda_{j}-\lambda_{j^{'}}+i\eta ) |} \nn \\
&\times& \frac{1}{\Pi_{m\beta l \neq n\alpha l^{'},l^{'}\pm1}  |\sin(\mu^{(n\alpha)}_{l}-\mu^{(m\beta)}_{l^{'}}+i\eta)   |}
~
\frac{|\det (H-2P)|^{2}}{|\det\Phi(\{\lambda\})|\cdot|\det \Phi^{r}(\{\mu\})|}.
\eea
\end{widetext}

\section{Deviation of string states}
\label{sect:deviation}

The string ansatz is known to be not exact even in the thermodynamic limit.
The solutions of rapidities may deviate from the pattern assumed by
string ansatz.
Such deviations must be taken into account when they are large
\cite{Hagemans2007}.
In this section, we give the formulae for an exact treatment of string
deviations for $1\chi^{(2)}R$ and $1\chi^{(3)}R$ excitations.

The branch cut of logarithmic function is taken as the negative real
axis which is identified with $\mathbb{R}^{-}+i0$.
From this the branch cut of $\arctan$-function is accordingly determined
via the definition
\bea
\arctan(z)=\frac{1}{2i}(\ln(1+iz)-\ln(1-iz)).
\eea

For a $1\chi^{(2)}R$ type excitation, let the two complex rapidities
be $\lambda^{(2)}_{\pm}=\lambda^{(2)}\pm i(\eta/2+\delta)$,
where $\delta$ represents the deviation from the pattern of string ansatz,
and the remaining $M-2$ real rapidities be $\{\lambda_{k}\}_{1\leq k \leq M-2}$.
Let the corresponding BQN be $J_{\pm}$ and $\{J_{k}\}_{1\leq k \leq M-2}$.
Then the two BAE for the complex rapidities are
\bea
N\theta_{1}(\lambda^{(2)}_{a})=&2\pi J_{a} +  \theta_{2}(\lambda^{(2)}_{a}-\lambda^{(2)}_{-a}) \nn\\
&\,\,\,+ \sum_{k=1}^{M-2} \theta_{2} (\lambda^{(2)}_{a}-\lambda_{k}),
\label{eq:2string}
\eea
where $a=\pm$.
In the followings, we assume that $\lambda^{(2)}\neq0$, $\delta \neq 0$, and $\lambda^{(2)}-\lambda_{j}\neq 0$, $1\leq j \leq M-2$.

From the choice of branch cut for $\arctan$-function,
the real part of the difference between the equations of $a=+$ and $a=-$ in Eq. (\ref{eq:2string}) gives
\bea
J_{-}-J_{+}=\Theta(\delta),
\eea
in which $\Theta (x)=1$ when $x \geq 0$, and $\Theta(x)=0$ when $x<0$.
Taking the sum of the equations for $a=+$ and $a=-$ in Eq. (\ref{eq:2string}), setting $\delta=0$, and comparing with the reduced BGT equation, we obtain
\bea
J_{-}+J_{+}=I^{(2)}+N\floor{\frac{\lambda^{(2)}}{\pi}+\frac{1}{2}} +\frac{N}{2} (-)^{\floor{\frac{\lambda^{(2)}}{\pi/2}}}.
\label{eq:two string sum}
\eea
The sign of $\delta$ can be determined from Eq. (\ref{eq:two string sum}) by noticing that $J_{\pm}$ are integers (half-integers) when $M$ is odd (even), i.e.
\bea
\Theta(\delta)=\mod(I^{(2)}-M+1+\frac{N}{2},2).
\label{eq:sign delta}
\eea
Combining Eqs.~(\ref{eq:two string sum},\ref{eq:sign delta}) together,
the BQN $J_{\pm}$ can be determined from the reduced one $I^{(2)}$ in BGT equations.
For the BQN of real rapidities, it can be shown that $J_{k}=I_{k}$, $1\leq k \leq M-2$.
To solve the exact values of rapidities, Eq. (\ref{eq:2string}) are replaced with the following two real equations.
The first one is the sum of the two equations in Eq. (\ref{eq:2string}), but not setting $\delta=0$.
The second one is obtained by taking the imaginary part of the $a=+$ equations in Eq. (\ref{eq:2string}), as
\bea
|\frac{\tan(\lambda^{(2)}_{+}-\lambda^{(2)}_{-})-i\tanh{\eta}}{\tan(\lambda^{(2)}_{+}-\lambda^{(2)}_{-})+i\tanh{\eta}}|=|\frac{\tan(\lambda^{(2)}_{+})-i\tanh{\eta/2}}{\tan(\lambda^{(2)}_{+})+i\tanh{\eta/2}}|^{N} \nn\\
\cdot\Pi_{k}| \frac{\tan(\lambda^{(2)}_{+}-\lambda_{k})+i\tanh{\eta}}{\tan(\lambda^{(2)}_{+}-\lambda_{k})-i\tanh{\eta}}|.\nn\\
\label{eq:delta deviation}
\eea
Combining these two equations with the BAE for real rapidities, the exact solutions can be solved.
The first order deviation of $\delta$ can be obtained from Eq. (\ref{eq:delta deviation}).
Up to first order of $\delta$, the left hand side (LHS) of Eq. (\ref{eq:delta deviation}) is ${|\delta|}/({\sinh(\eta)\cosh(\eta)}) $.

For the case of $1\chi^{(3)}R$ excitation, the logic is similar.
Let the three complex rapidities be $\lambda^{(3)}_{a}$ with $a=\pm,\,0$, and the real rapidities be  $\{\lambda_{k}\}_{1\leq k \leq M-3}$.
Let the corresponding Bethe quantum numbers be $J_{a}$ ($a=\pm,\, 0$), and $\{J_{k}\}_{1\leq k \leq M-3}$.
To parametrize the string deviations, the complex rapidities are written as $\lambda^{(3)}_{0}=\lambda^{(3)}$, and $\lambda^{(3)}_{\pm}=\lambda^{(3)}+\epsilon \pm i (\eta+\delta)$.
The BAE for the three complex rapidities are
\bea
N\theta_{1}(\lambda^{(3)}_{a})=&2\pi J_{a} +  \sum_{b\neq a} \theta_{2}(\lambda^{(3)}_{a}-\lambda^{(3)}_{b})  \nn\\
&\,\,\,+ \sum_{k=1}^{M-3} \theta_{2} (\lambda^{(3)}_{a}-\lambda_{k}),
\label{eq:3string}
\eea
where $a,\,b=\pm,\,0$.
We assume that $\lambda^{(3)}\neq 0$, $\epsilon\neq 0$, $\delta\neq 0$, and $\lambda^{(3)}-\lambda_{j}\neq 0$, $1\leq j \leq M-3$.

The real part of the difference between the equations for $a=+$ and $a=-$ in Eq.~(\ref{eq:3string}) gives
\bea
J_{-}-J_{+}=1.
\label{eq:three string difference}
\eea
Taking the sum of the three equations in Eq.~(\ref{eq:3string}), setting $\epsilon=\delta=0$, and comparing with the reduced BGT equation, we obtain
\bea
J_{+}+J_{0}+J_{-}=&I^{(3)}+N(2\floor{\frac{\lambda^{(3)}}{\pi}+\frac{1}{2}}+(-)^{\floor{\frac{\lambda^{(3)}}{\pi/2}}} )\nn\\
&-\sum_{k} (\floor{\frac{\lambda^{(3)}-\lambda_{k}}{\pi}+ \frac{1}{2}} +\frac{1}{2} (-)^{\floor{\frac{\lambda^{(3)}-\lambda_k}{\pi/2}} }  ). \nn \\
\label{eq:3string bethe num sum}
\eea
To determine $J_{\pm}$ and $J_0$, the sum of the equations for $a=\pm$ in Eq. (\ref{eq:3string}) is taken, yielding
\begin{widetext}
\be
2\pi(J_{+}+J_{-}) +\theta_{2}(\lambda^{(3)}_{+}-\lambda^{(3)}_{0})+\theta_{2}(\lambda^{(3)}_{-}-\lambda^{(3)}_{0})=
N(\theta_{1}(\lambda^{(3)}_{+})+\theta_{1}(\lambda^{(3)}_{-}))-\sum_{k} (\theta_{2}(\lambda^{(3)}_{+}-\lambda_{k}) +\theta_{2}(\lambda^{(3)}_{-}-\lambda_{k})).\\
\label{eq: three string sum pm}
\ee
\end{widetext}
Define A to be the right hand side of Eq.~(\ref{eq: three string sum pm}).
Since $\theta_{2}(\lambda^{(3)}_{+}-\lambda_{0})+\theta_{2}(\lambda^{(3)}_{-}-\lambda_{0})\in (-2\pi,2\pi)$,
$J_{+}+J_{-}$ is the even (odd) integer number within $(A/2\pi-1,A/2\pi+1)$ when $M$ is even (odd).
Hence
\bea
&J_{+}+J_{-}=(1+(-)^M) \floor{\frac{1}{2} (\frac{A}{2\pi}+1)}\nn\\
&\,\,\,\,\,\,\,\,\,\,\,\,\,\,\,\,\,\,\,\,\,\,\,\,\,\,\,\,\,\,\,\,\,\,\,\,\,+(1-(-)^M) (\floor{\frac{1}{2} (\frac{A}{2\pi} +1)+\frac{1}{2} }-\frac{1}{2} ).\nn\\
\label{eq:3string bethe num sum pm}
\eea
From Eqs. (\ref{eq:three string difference}, \ref{eq:3string bethe num sum}, \ref{eq:3string bethe num sum pm}),
the values of $J_{\pm}$ and $J_0$ can be determined from the reduced BQN $I^{(3)}$ in BGT equation.
The BQN for real rapidities can be proved to be of the following expression in similar manner,
\bea
J_k=I_k-\floor{\frac{\lambda_k-\lambda^{(3)}}{\pi}+\frac{1}{2} } - \frac{1}{2} (-)^{\floor{\frac{\lambda_k-\lambda^{(3)}}{\pi/2} } },
\label{eq: 3string bethe num real}
\eea
where $1\leq k \leq M-3$.

For solving rapidities, Eq. (\ref{eq:3string}) are replaced with the following three real equations.
The first one is the sum of the equations for $a=\pm$, $a=0$ in Eq. (\ref{eq:3string}) without setting $\epsilon$ and $\delta$ to be zero.
The second one is Eq. (\ref{eq: three string sum pm}).
The third one is by taking imaginary part of the difference between the equations for $a=+$ and $a=-$ in Eq. (\ref{eq: three string sum pm}), which is
\begin{widetext}
\bea
|\frac{\tan(\lambda^{(3)}_{+}-\lambda^{(3)}_{0})-i\tanh{\eta}}{\tan(\lambda^{(3)}_{+}-\lambda^{(3)}_{0})+i\tanh{\eta}}|=|\frac{\tan(\lambda^{(3)}_{+}-\lambda^{(3)}_{-})+i\tanh{\eta}}{\tan(\lambda^{(3)}_{+}-\lambda^{(3)}_{-})-i\tanh{\eta}}|
\cdot |\frac{\tan(\lambda^{(3)}_{+})-i\tanh{\eta/2}}{\tan(\lambda^{(3)}_{+})+i\tanh{\eta/2}}|^{N} \cdot\Pi_{k} |\frac{\tan(\lambda^{(2)}_{+}-\lambda_{k})+i\tanh{\eta}}{\tan(\lambda^{(2)}_{+}-\lambda_{k})-i\tanh{\eta}}|. \nn\\
\label{eq:three string difference imag}
\eea
\end{widetext}
Let $\epsilon=r\sin\theta$, $\delta=r\cos\theta$.
For first order deviation, we remark that up to first order in $\epsilon$ and $\delta$, the LHS of Eq. (\ref{eq:three string difference imag}) is ${r}/({2\sinh \eta \cosh \eta})$,
and $\theta$ can be determined from Eq. (\ref{eq: three string sum pm}) as
\bea
\theta=-\phi+\pi \text{sign} \phi,
\eea
in which $\phi$ is defined to be $\frac{1}{2}A-\pi J_{0}$.
The values of $r$ and $\theta$ can be used as the initial inputs in an iterative solution of $\epsilon$ and $\delta$.

\section{Sum rules}
\label{sect:sum}

The momentum-resolved first frequency sum rules are presented below.
The transverse first frequency moment (FFM) sum rule is
$
W_\perp(q) = \int_{0}^\infty  \frac{d \omega}{2\pi}~
\omega \left[ {S^{ +  -} (q,\omega ) + S^{ -  + } (q,\omega )} \right]
=   \alpha_\perp +\beta_\perp \cos q ,
$
where $\alpha_\perp= -e_0-\Delta{\partial e_0}/{\partial \Delta}+mh$
and $\beta_\perp =(2-\Delta ^2){\partial e_0}/{\partial \Delta}
+\Delta e_0$.
Its longitudinal version is also known as $W_\pp(q) =
\int_{0}^\infty \frac{d \omega}{2\pi} \omega S^{zz} (q,\omega )
=(1-\cos q) \alpha_\pp$ \cite{Mossel2008},
where $\alpha_2=-e_0+\Delta {\partial e_0}/{\partial \Delta}$.

Here we summarize the derivation of the first frequency moment
sum rule in Eq. (\ref{eq:ffm}) following Ref. [\onlinecite{Mossel2008}].
The first frequency moment is defined as
\bea
\omega_{a \bar{a}} (q)=\int_{-\infty}^{\infty} \frac{d\omega}{2\pi}
\omega S^{a\bar{a}} (q,\omega).
\label{eq:first sum def}
\eea
The expressions of $\omega_{+-}+\omega_{-+}$ and $\omega_{zz}$ are derived
as a function of $\Delta$ and $h$ for the XXZ Hamiltonian (Eq.(1) in main text).

By inserting a complete set of eigenstates and performing the integration
with respect to $t$ and $\omega$, $\omega_{ii}$ ($i=x,y,z$) can be
 transformed as
\begin{widetext}
\be
\omega_{ii}=\frac{1}{N} \sum_{j,j^{'}} e^{-i q (j-j^{'})}
\int_{-\infty}^{\infty} \frac{d\omega}{2\pi} \int_{-\infty}^{\infty}
dt \omega e^{i \omega t}  \sum_{\mu} e^{i(E_{G}-E_{\mu})t} \bra{G} S^{i}_{j} \ket{\mu} \bra{\mu} S^{i}_{j^{'}} \ket{G} \nn
= -\frac{1}{N} \sum_{j,j^{'}} e^{-i q (j-j^{'})}
\bra{G} [H,S^{a}_{j}] S^{a}_{j^{'}}  \ket{G}.
\ee
\end{widetext}
Similarly
\bea
\omega_{ii}=\frac{1}{N} \sum_{j,j^{'}} e^{-i q (j-j^{'})}  \bra{G}
S^{i}_{j} [H,S^{i}_{j^{'}}] \ket{G}.
\label{eq:first sum eq2}
\eea
Since the system is invariant under inversion tranformation defined as $P\vec{S_{j}}P^{-1}=\vec{S}_{-j}$, i.e.
\bea
P\ket{G} = \ket{G}, \,\,\,\,\, PHP^{-1}=H,
\eea
Eq. (\ref{eq:first sum eq2}) becomes
\bea
\omega_{ii}
&=& \frac{1}{N} \sum_{j,j^{'}} e^{-i q (j-j^{'})}
\bra{G} S^{i}_{j^{'}} [H,S^{i}_{j}] \ket{G},
\eea
where in obtaining the last line the change of summation indices
$-j \rightarrow j^{'}$ and $-j^{'} \rightarrow j$ is performed.
Combining these results together, we obtain
\bea
\omega_{ii}&=& -\frac{1}{2N} \sum_{j,j^{'}} e^{-i q (j-j^{'})}
\bra{G} [[H, S^{i}_{j}], S^{i}_{j^{'}}] \ket{G},
\eea
The commutation relations for $i=x,y,z$ can be carried out explicitly,
and the results for $\omega_{ii}$ are
\begin{widetext}
\bea
\omega_{xx(yy)}&=&-\frac{1}{N} \sum_{j} [ (1-\Delta \cos q)
  \bra{G} S^{y(x)}_{j} S^{y(x)}_{j+1}\ket{G}
+(\Delta-\cos q) \bra{G} S^{z}_{j} S^{z}_{j+1} \ket{G}-\frac{h}{2} S^{z}_{j} ],
\nn \\
\omega_{zz}&=&-\frac{1}{N}(1-\cos q)\sum_j \bra{G}
(S^{x}_{j} S^{x}_{j+1}+ S^{y}_{j} S^{y}_{j+1})\ket{G}.\nn\\
\eea
\end{widetext}

In the main text $S^{+-}(q,\omega)$ and $S^{-+}(q,\omega)$ are calculated,
and their first frequency moment sum rule can be derived
from $\omega_{xx}$ and $\omega_{yy}$ through
\bea
\omega_{+-}+\omega_{-+}=2(\omega_{xx}+\omega_{yy}).
\eea
Under the help of the Hellman-Feynman theorem, we have
\bea
\bra{G} \sum_{j} S^{z}_{j} S^{z}_{j+1}
\ket{G}&=& \frac{\partial e_{0}}{\partial \Delta}, \nn \\
\bra{G} \sum_{j} (S^{x}_{j} S^{x}_{j+1}+S^{y}_{j} S^{y}_{j+1})
\ket{G}&=& e_{0}- \Delta \frac{\partial e_{0}}{\partial \Delta}. \nn \\
\eea
where $e_0$ is defined as
\bea
e_{0}= \sum_{j} \bra{G}  (S^{x}_{j} S^{x}_{j+1}+S^{y}_{j} S^{y}_{j+1} +\Delta S^{z}_{j} S^{z}_{j+1}) \ket{G}.
\eea
The magnetic field $h$ and magnetization $m$ are related through
the Legendre transform
\bea
h=\frac{1}{N} \frac{\partial e_{0}}{\partial m}.
\eea
Combining these results together, the first frequency moment sum
rule can be expressed as
\bea
\omega_{+-}(q) + \omega_{-+}(q)&=&-\frac{2}{N}
[ (\Delta(1+\Delta \cos q)-2 \cos q) \frac{\partial e_{0}}
{\partial \Delta} \nn \\
&+& (1-\Delta \cos q) e_{0} -m \frac{\partial e_{0}}{\partial m} ],
\eea
\bea
\omega_{zz}(q)=-\frac{1}{N} (1-\cos q)
(e_0-\Delta \frac{\partial e_0}{\partial \Delta}).
\eea

\section{Bethe eigenstates at spectral peak positions in transverse DSFs}
\label{sect:state_peak}

In this section, we identify the Bethe eigenstates with the largest weight values around the spectral peaks at momenta $0,\frac{\pi}{2},\pi$.
The energies of these eigenstates can be obtained by solving the Bethe
ansatz equations, which correspond to the peak positions in the DSSF
spectra as shown in Fig. \ref{fig:bright_line}.
In the following, $S^z_T=\sum_{i=1}^N S^z$ is the $z$-component of
the total spin; $M=\frac{N}{2}-S^z_T$ is the number of magnons;
and $m=S^z_T/N$ is the magnetization per site.
For simplicity, we assume that both $N$ and $S^z_T$ are even integer numbers.
For the expressions of the momentum $k$ of the excitations $\chi^{(n)}$ ($n=1,2$), $\psi$ and $\psi^*$,
the limit of $N\rightarrow \infty$ is taken with $m$ fixed.


\begin{figure}
\centering\epsfig{file=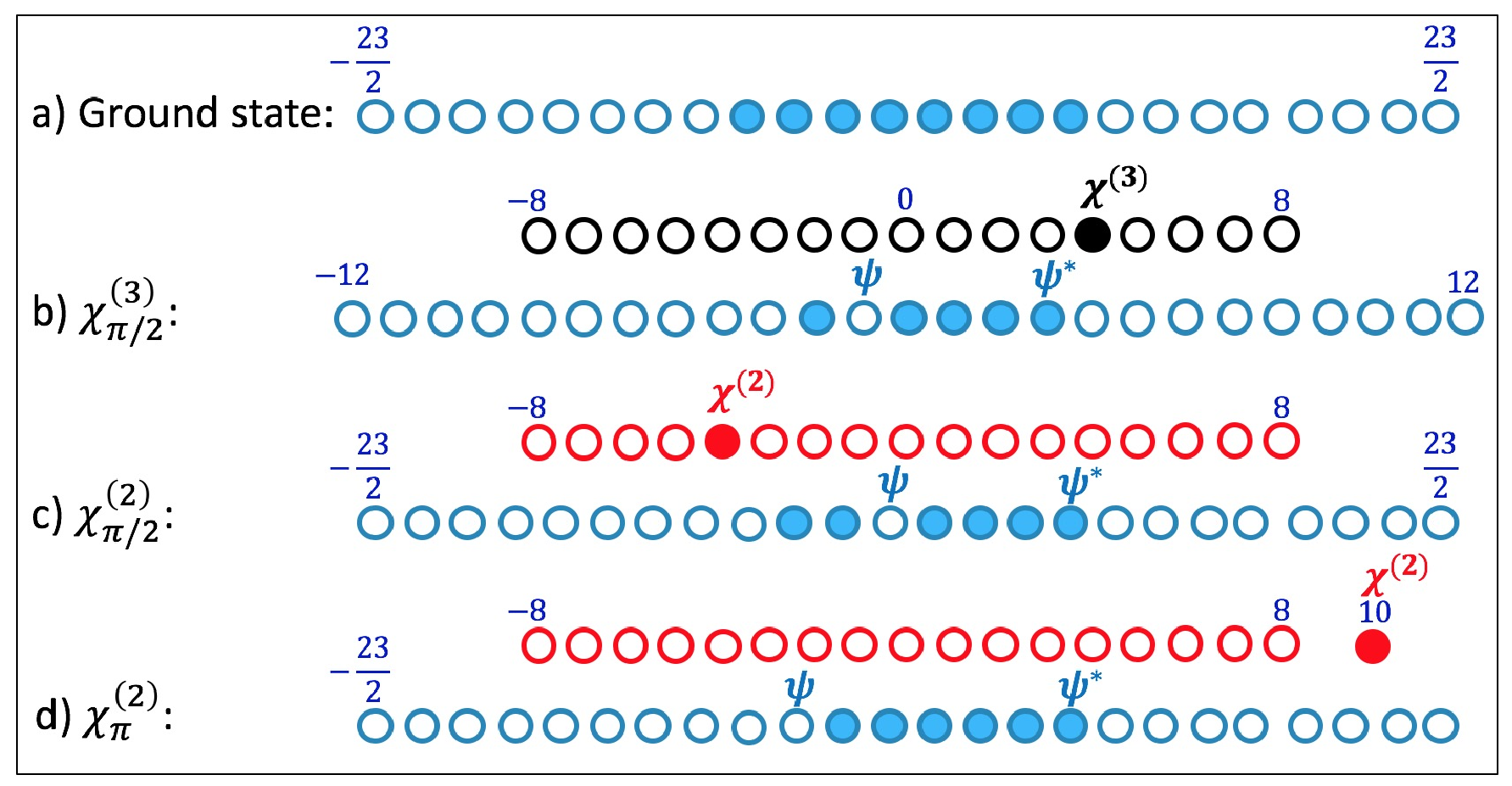,clip=1.5,width=1.0\linewidth,angle=0}
\caption{Distributions of Bethe quantum numbers for the string excitations which have local maximal weight values at the corresponding momentum.
The positions of the solid circles represent the Bethe quantum numbers of the particles.
The system size and magnetization are taken as $N=32$ and $S^z_T=8$.
}
\label{fig:states_string}
\end{figure}

For the line of  $\chi^{(3)}_{\frac{\pi}{2}}$ in Fig. \ref{fig:bright_line},
the Bethe quantum numbers of the corresponding Bethe eigenstate are given by
\bea
I^{(3)}&=&\frac{1}{2}S^z_T, \nn \\
I^{(1)}_j&=&-\frac{M-4}{2}+j-1+\Theta(j-\frac{M}{2}+3),
\eea
where $1\leq j \leq M-3$, and $\Theta$ is the step function defined as $\Theta(x)=0$ if $x\leq 0$ and $\Theta(x)=1$ if $x> 0$.
The momenta of the excitations are determined by Eq. (\ref{eq:string_momentum}) as $k_{\chi^{(3)}}=\pi(1-m)$, $k_{\psi}=0$ and $k_{\psi^*}=\pi (1/2+m)$.

For the line of  $\chi^{(2)}_{\pi/2}$, the Bethe quantum numbers of the corresponding Bethe eigenstate are
\bea
I^{(2)}&=&-\frac{1}{2}S^z_T,\nn \\
I^{(1)}_j&=&-\frac{M-3}{2}+j-2+\Theta(j-\frac{M}{2}+1),
\eea
where $1\leq j \leq M-2$.
The momenta of the excitations are $k_{\chi^{(2)}}=\pi(1+m)$, $k_{\psi}=0$ and $k_{\psi^*}=\pi(3/2-m)$.

For the line of  $\chi^{(2)}_{\pi}$, the Bethe quantum numbers of the corresponding Bethe eigenstate are
\bea
I^{(2)}&=&S^z_T+2,\nn \\
I^{(1)}_j&=&-\frac{M-3}{2}+j,\nn\\
\eea
where $1\leq j \leq M-2$.
The momenta of the excitations are $k_{\chi^{(2)}}=\pi(1-2m)$, $k_{\psi}=k_{\psi^*}=\pi(1/2+m)$.

For the line of  $R^{-+}_{\pi/2}$ ($m\leq1/4$), the Bethe quantum numbers of the corresponding Bethe eigenstate are
\bea
I^{(1)}_j=-\frac{M-1}{2}+j-1+\Theta(j-M+\frac{N}{4}),
\eea
where $1\leq j \leq M$.
The momenta of the excitations are $k_{\psi_1}=\pi(1/2+m)$ and $k_{\psi_2}=\pi(1-m)$.

For the line of  $R^{+-}_{0}$, the Bethe quantum numbers of the corresponding Bethe eigenstate are
\bea
I^{(1)}_j&=&-\frac{M-1}{2}+j,1\leq j\leq M-1,\nn\\
I^{(1)}_M&=&\frac{M-1}{2}+S^z_T+1.
\eea
The momenta of the excitations are $k_{\psi}=\pi(1/2+m)$ and $k_{\psi^*}=\pi(1/2-m)$.

For the line of  $R^{+-,a}_{\pi/2}$, the Bethe quantum numbers of the corresponding Bethe eigenstate are
\bea
I^{(1)}_j&=&-\frac{M-1}{2}+j,1\leq j\leq M-1,\nn\\
I^{(1)}_M&=&\frac{N}{4}-\frac{M-1}{2}.
\eea
The momenta of the excitations are $k_{\psi}=\pi(1/2+m)$ and $k_{\psi^*}=\pi(1-m)$.

For the line of  $R^{+-,b}_{\pi/2}$, the Bethe quantum numbers of the corresponding Bethe eigenstate are
\bea
I^{(1)}_j&=&-\frac{M-1}{2}+j,1\leq j\leq M-1,\nn\\
I^{(1)}_M&=&\frac{N}{4}+\frac{M-1}{2}.
\eea
The momenta of the excitations are $k_{\psi}=\pi(3/2-m)$ and $k_{\psi^*}=\pi m$.

Schematically, we present the distributions of Bethe quantum
numbers of string excitations are shown in Fig. \ref{fig:states_string}.



\end{document}